\pdfoutput=1
\documentclass[sigconf]{acmart}
\usepackage{hyperref}
\usepackage{balance}

\setcopyright{rightsretained}
\acmPrice{15.00}

\setcopyright{usgovmixed}
\acmDOI{0000001.0000001}

\usepackage{graphicx}
\usepackage{subcaption}
\usepackage{pgfplots}
\usepackage{epstopdf}

\newcommand{\task}{218}
\newcommand{\company}{BigCompany}

\begin{document}

%% Title information
\title[]{Crowdtesting : When is The Party Over?}
\author{Junjie Wang$^1$, Ye Yang$^2$, Zhe Yu$^3$, Tim Menzies$^3$, Qing Wang$^1$}
% \authornote{Junjie Wang, Ye Yang, Zhe Yu, Tim Menzies, Qing Wang}          %% \authornote is optional;
                                        %% can be repeated if necessary
% \orcid{nnnn-nnnn-nnnn-nnnn}             %% \orcid is optional
\affiliation{
  \position{$^1$Institute of Software Chinese Academy of Sciences, Beijing, China}
  \department{$^2$School of Systems and Enterprises, Stevens Institute of Technology, USA}    
  \institution{$^3$Department of Computer Science, North Carolina State University, Raleigh, NC, USA}            %% \institution is required
  \country{ }                    %% \country is recommended
}
\email{{wangjunjie,wq}@itechs.iscas.ac.cn, ye.yang@stevens.edu, zyu9@ncsu.edu, tim@menzies.us}          %% \email is recommended

%% Author with two affiliations and emails.
% \author{First2 Last2}
% \authornote{with author2 note}          %% \authornote is optional;
%                                         %% can be repeated if necessary
% \orcid{nnnn-nnnn-nnnn-nnnn}             %% \orcid is optional
% \affiliation{
%   \position{Position2a}
%   \department{Department2a}             %% \department is recommended
%   \institution{Institution2a}           %% \institution is required
%   \streetaddress{Street2a Address2a}
%   \city{City2a}
%   \state{State2a}
%   \postcode{Post-Code2a}
%   \country{Country2a}                   %% \country is recommended
% }
% \email{first2.last2@inst2a.com}         %% \email is recommended
% \affiliation{
%   \position{Position2b}
%   \department{Department2b}             %% \department is recommended
%   \institution{Institution2b}           %% \institution is required
%   \streetaddress{Street3b Address2b}
%   \city{City2b}
%   \state{State2b}
%   \postcode{Post-Code2b}
%   \country{Country2b}                   %% \country is recommended
% }
% \email{first2.last2@inst2b.org}         %% \email is recommended

%% Abstract
%% Note: \begin{abstract}...\end{abstract} environment must come
%% before \maketitle command
\begin{abstract}
Trade-offs such as ``how much testing is enough'' are critical yet challenging project decisions in software engineering. Most existing approaches adopt risk-driven or value-based analysis to prioritize test cases and minimize test runs. 
However, none of these is applicable to the emerging crowd testing paradigm where task requesters typically have no control over online crowdworkers's dynamic behavior and uncertain performance.

% and deciding when to close a crowdtesting task is largely done by guesswork.
% crowdworkers are encouraged to come and perform the testing tasks at any time and could not be prioritized.
% testing tasks are decomposed into many micro-tasks to enable massive parallel testing performed by unknown online testers.

In current practice, deciding when to close a crowdtesting task is largely done by guesswork due to lack of decision support. 
This paper intends to fill this gap by introducing automated decision support for monitoring and determining appropriate time to close the crowdtesting tasks.

First, this paper investigates the necessity and feasibility of close prediction of crowdtesting tasks based on industrial dataset. Then, it designs 8 methods for close prediction, based on various models including the bug trend, bug arrival model, capture-recapture model. 
Finally, the evaluation is conducted on 218 crowdtesting tasks from one of the largest crowdtesting platforms in China, and the results show that a median of 91\% bugs can be detected with 49\% saved cost.

% To bridge the gap,  this paper investigates the close prediction problem for crowdtesting tasks and designs automated methods based on a variety of classical defect arrival models. It starts with a motivational study to demonstrate the necessity and feasibility of close prediction for crowdtesting tasks, then it formulates the close prediction problem , designs 8 automated methods based different models including Rayleigh model, Knee model, Capture-ReCapture model and its variants, and  
% First, this paper . 
% Then, it designs 8 methods for close prediction, based on various models including the bug trend curve, bug arrival model, capture-recapture model.
% Finally, the evaluation is conducted on 218 crowdtesting tasks from one of the largest crowdtesting platforms in China, and the results show that a median of 91\% bugs can be detected with 49\% saved cost.

%The most straightforward method produces the most effective and stable performance.
\end{abstract}

%% 2012 ACM Computing Classification System (CSS) concepts
%% Generate at 'http://dl.acm.org/ccs/ccs.cfm'.
% <concept_id>10011007.10011006.10011008</concept_id>
% <concept_desc>Software and its engineering~General programming languages</concept_desc>
% <concept_id>10003456.10003457.10003521.10003525</concept_id>
% <concept_desc>Social and professional topics~History of programming languages</concept_desc>
% <concept_significance>300</concept_significance>

% \ccsdesc[500]{Software and its engineering~Software testing
% and debugging}
% \ccsdesc[300]{Software and its engineering~Collaboration
% in software development}

%% End of generated code

%% Keywords
%% comma separated list
  %% \keywords are mandatory in final camera-ready submission

%% \maketitle
%% Note: \maketitle command must come after title commands, author
%% commands, abstract environment, Computing Classification System
%% environment and commands, and keywords command.
\maketitle

\section{Introduction}
\label{sec:intro}
Crowdtesting is an emerging trend in software testing practices that accelerates testing processes by attracting online crowd workers to accomplish various types of testing tasks \cite{mao2017survey,chen2018research,link_crowdtest,wang2017domain,cui2017who}. On one hand, crowdtesting entrusts testing tasks to unknown, online crowd workers whose diverse testing environments/platforms, background, and skill sets could significantly contribute to more reliable, cost-effective, and efficient testing results. On the other hand, some aspects of software cannot be tested any other way, except asking humans to use the system, e.g., usability testing \cite{gomide2014affective}. For these tasks, crowdtesting is inherently a natural fit than other alternatives. 

Trade-offs such as ``how much testing is enough'' are critical yet challenging project decisions in software engineering \cite{myers2011art,lewis2016software,garg2011stop,iqbal2013software}.
Stopping too early can lead to inefficient testing and unsatisfying software quality, while stopping too late can result in the waste of testing resources.
% The crowdtesting tasks also face with such concern \cite{mao2017survey,chen2018research}.

In current practice, deciding when to close a crowdtesting task is largely done by guesswork due to lack of decision support. % for most crowdtesting platforms.
Project managers usually set up task's closing condition through either a fixed duration (e.g., 5 days) or fixed budget (e.g., recruiting 400 crowd workers).
If either of the criteria is met first, then the task will be automatically closed.
% To ensure sufficient testing, the tester tends to employ a relatively large threshold for testing period or the budget, which leads to a waste of resources.
Our investigation on real-world crowdtesting data reveals that the number of detected bugs\footnote{This paper uses \textit{bug} and \textit{defect} interchangeably.} would first increase rapidly, then undergo slow growth, and finally become flatten-out (see Section \ref{subsec:background_observations}).
This is because for the latter stage of a crowdtesting task, the submitted reports are mainly the duplicates of previous ones.
Therefore, it is of great value to automatically decide when to close a crowdtesting task so as to improve its cost-effectiveness.

% Intuitively, this observation is consistent with the defect arrival Rayleigh model, which indicates that there might exist an optimal time to close the crowdtesting tasks, and after that point, the submitted test reports are mainly the duplicates of previously submitted ones. The current crowdtesting practices generally overlooked this perspectives, which would result in unnecessary waste of testing resources.}

Many existing approaches employed either risk-driven or value-based analysis to prioritize or select test cases and minimize test runs \cite{wang2017qtep,shi2015comparing,harman2015empirical,saha2015information,henard2016comparing}. 
However, none of these is applicable to the emerging crowd testing paradigm where task requesters typically have no control over online crowdworkers's dynamic behavior and uncertain performance.
There were several researches focusing on the time-series models for measuring software reliability and predicting when to stop testing and release a software product \cite{garg2011stop,garg2013method,iqbal2013software}.
This paper will adopt two most promising models (i.e., Rayleigh's defect arriving model and knee model) and examine its effectiveness in predicting when to close a crowdtesting task.
Another body of previous researches aimed at optimizing software inspection, which also concerned predicting the total and remaining number of bugs  \cite{briand2000comprehensive,walia2009evaluating,chun2006estimating,rong2017towards,mandala2012application,goswami2015using,vitharana2017defect}. 
This paper will adopt ideas from the most commonly-used capture-recapture models and examine its effectiveness in conducting close prediction of crowdtesting.

% Since there is no existing work for close prediction of crowdtesting task, 
% To explore the feasibility of close prediction of crowdtesting task, 
This paper first investigates the necessity and feasibility of close prediction of crowdtesting tasks based on industrial dataset.
It then designs 8 methods to conduct the close prediction.
Method \textit{Trend} is a straightforward and intuitive method, which determines the close time if no new bugs detected over a certain number of successive reports. 
% Method \textit{Trend} is a straightforward and intuitive method, which decides the close if the number of reported bugs remains unchanged for a certain number of successive reports.
Method \textit{Peak} is based on the Rayleigh's defect arriving model, while method \textit{Knee} is based on the slope of bug trend curve.
The other five methods are based on different variations of Capture-ReCapture models, which can estimate the total number of bugs in a software system.
% When the detected number of bugs so far equals to the estimated total number of bugs, the methods would predict the time, when the last report was received, as the close time. 

This paper evaluates each method on {\task} crowdtesting tasks from one of the largest Chinese crowdtesting platforms.
The experimental results show that the most straightforward \textit{Trend} method achieves the best performance.
% in terms of effectiveness and stability, 
Generally speaking, a median of 91\% bugs can be detected with 49\% reduced cost.
For our experimental crowdtesting platform, it delivers about 1000 crowdtesting tasks a year and a task consumes approximately 3,000 China Yuan (i.e., the cost paid to crowdworkers).
According to estimates, this crowdtesting platform can save 1,470,000 China Yuan (about 245,000 US dollars) a year.

The contributions of this paper are as follows:

\begin{itemize}
\item An empirical investigation on the necessity and feasibility to conduct the close prediction of crowdtesting tasks based on industrail dataset. To the best of our knowledge, this is the first work to identify the problem of close prediction for crowdtesting task. Practical experiences lead us to believe that this is an important problem in crowdtesting.

\item The design of 8 methods for close prediction of crowdtesting tasks.
% , and experimentally recommend the best method for close prediction.

\item An evaluation of the effectiveness of the designed methods based on {\task} crowdtesting tasks from one of the largest crowdtesting platforms in China, and results are promising.

\item A cautionary tale that verbatim reuse of methods from other fields may not produce the best results of crowdtesting. Specifically, we show the capture-recapture models from software inspections do not work well on crowdtesting data. Furthermore, a straightforward method can produce the most effective performance in close prediction of crowdtesting task\footnote{Url for the website with experimental dataset, source
code and detailed experimental results is blinded for review.}.

\end{itemize}

Note that, this paper does not aims at exploring the entire set of methods for close prediction of crowdtesting. 
Instead, because there are no ready-made methods, we adopt ideas from several commonly-used and representative techniques\cite{lyu1996handbook,yamada2014software}, and design 8 methods for close prediction of crowdtesting, with demonstrated prediction capability as well as application readiness.

The rest of this paper is organized as follows. Section \ref{sec:background} describes the background and motivation of this study. Section \ref{sec:approach} presents the details of our designed methods. Sections \ref{sec:experiment} and \ref{sec:result} show the experimental setup and evaluation results
respectively. Section \ref{sec:discussion} provides a detailed discussion and threats to validity. Section \ref{sec:related} surveys related
work. Finally, we summarize this paper in Section \ref{sec:conclusion}.

\section{Background and Motivation}
\label{sec:background}

\subsection{Background}
\label{subsec:background_crowdtesting}

In this section, we present a brief background of crowdtesting to help better understand the challenges we meet in real industrial crowdtesting practice.

\begin{figure}[t!]
\centering
\vspace{0.1in}
\includegraphics[width=5.5cm]{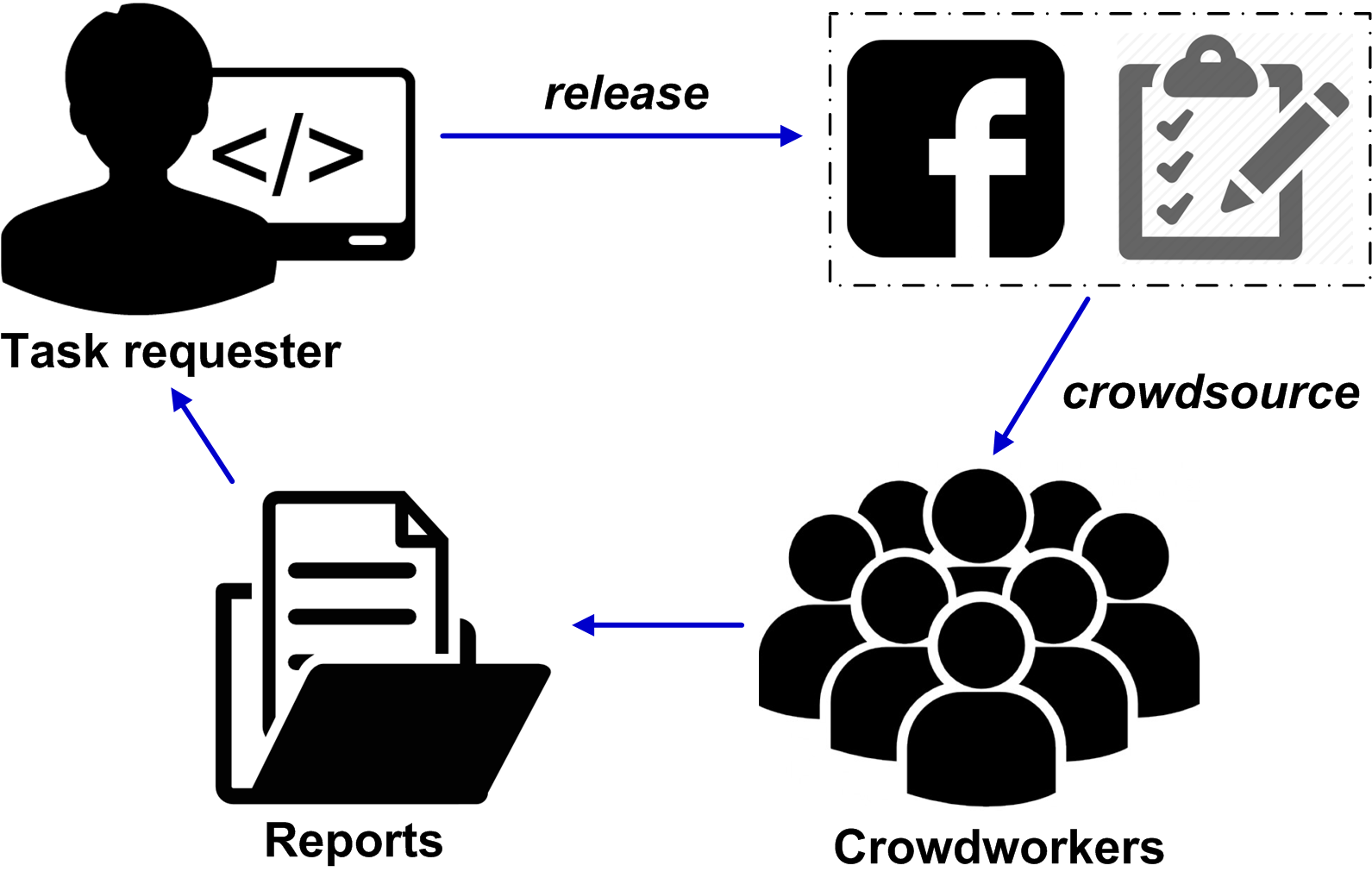}
\caption{Procedure of crowdtesting \cite{feng2015test}}
\label{fig:process}
% \vspace{-0.1in}
\end{figure}

As shown in Figure \ref{fig:process}, in general, the task requester prepares the crowdtesting task (including the software under test and test requirements), and distributes it on the crowdtesting platform.
The crowdworkers can sign in to conduct the tasks and are required to submit crowdtesting reports, which describe the input, operation steps, results of the test, etc.
% Here, we will describe the context of crowdtesting process shared with these methods.
In this way, the crowdtesting platform will receive crowdtesting reports submitted by the crowdworkers in chronological order. 
The task requester then inspects each report manually or using automatic tool support (e.g., \cite{wang2016towards,wang2017domain}), and the content of each report will be characterized using two attributes: 1) whether it contains a bug\footnote{In our experimental platform, a report would contain zero or one bug.}; 2) whether it is the duplicate of previously submitted reports. 
In the following paper, if not specified, when we say ``\textit{bug}'' or ``\textit{unique bug}'', we mean the corresponding report contains a bug and the bug is not the duplicate of previously submitted ones. 
% we treat a report as a bug only when it contains bug and the bug is not the duplicate of previously submitted ones. 

Our experiment is conducted with {\company} crowd-testing platform.
Through talking with the project managers in this platform, we find that deciding when to close a crowdtesting task is largely done by guesswork due to lack of decision support. 
% deciding when to close a crowdtesting task is kind of guesswork in current practice.
They usually set up either a fixed period (e.g., 5 days) or a fixed number of participant (e.g., recruiting 400 crowd workers).
If either of the criteria is met first, then the testing task will be automatically closed.
To avoid insufficient testing, they tend to employ a relatively large threshold for testing period or number of participants.
The observation on their dataset reveals it is a waste of cost (see Section \ref{subsec:background_observations}).
The project managers mentioned that they keen for the automatic decision support for when to close a crowdtesting task.

\subsection{{\company} DataSet}
\label{subsec:background_dataset}

The experimental dataset is collected from {\company}\footnote{Blinded for review.} crowdtesting platform, which is one of the largest crowdtesting platforms in China.
%Baidu CrowdTest was founded in 2011 and has become one of the largest crowdsourced testing platforms in China.
We collected all crowdtesting tasks closed between May. 1st 2017 and Jul. 1st 2017.
In total, there are 218 crowdtesting tasks, with 46434 submitted crowdtesting reports.
The minimum, average, and maximum number of reports (\textit{and unique bugs}) in a crowdtesting task are respectively 101 (\textit{6}), 213 (\textit{26}), and 876 (\textit{89}).
% The minimum, average, and maximum number of bugs in a crowdtesting task are respectively 6, 26, and 89.

To understand the real-world crowdtesting practice, we have conducted an analysis on the collected dataset, and observations are shown in the next subsection.

\subsection{Observations and Implications}
\label{subsec:background_observations}

For the received reports (in chronological order) of each crowdtesting task, we  count how many unique bugs have been accumulated considering the previous \textit{K} reports (we call it \textit{bug trend} for simplicity).
% each crowdtesting task, we first rank its submitted reports in chronological order, and
\textit{K} is ranged from 1 to the total number of reports.
We then compute the percentage of bugs for each \textit{K}, considering the total number of detected unique bugs.
% Note that, all duplicate bugs from different reports will be treated as one (unique) bug.

We have investigated the bug trend for all {\task} experimental crowdtesting tasks.
A general pattern observed is that the number of detected bugs of a test task would first increase rapidly, then undergo slow growth, and finally become flatten-out.
This is because for the latter stage of a crowdtesting task, the submitted reports are mainly contributing duplicate bugs.

\begin{figure}[t!]
\centering
\includegraphics[width=9cm]{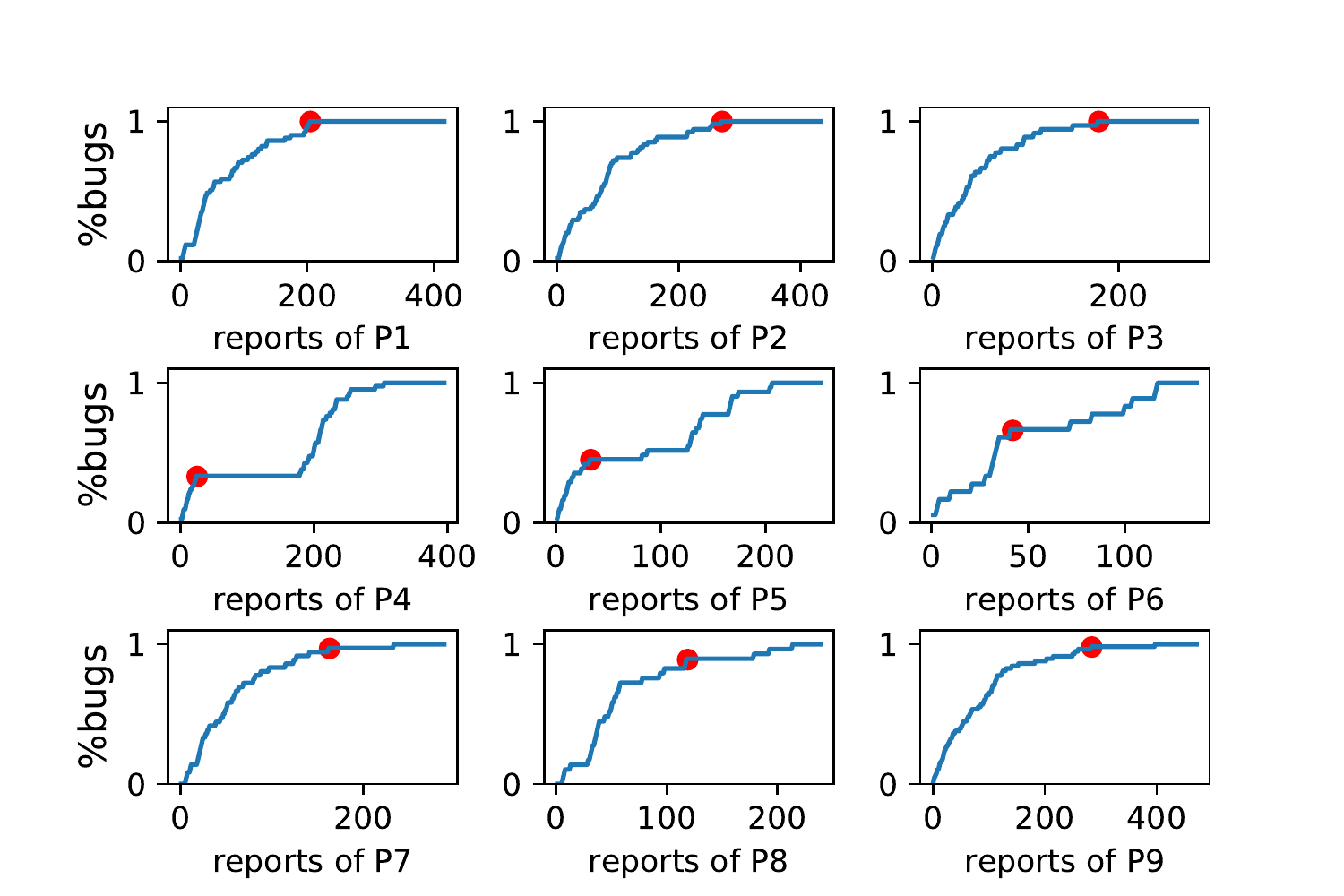}
\caption{Observation of bug trend based on BigCompany dataset}

\label{fig:observation}
% \vspace{-0.1in}
\end{figure}

Nevertheless, as the crowdworkers are encouraged to come and perform the testing tasks at any time, the bug trend of different tasks vary slightly. 
We further summarize three categories of bug trend in Figure \ref{fig:observation} to better motivate this study. 
The red dots in Figure \ref{fig:observation} denotes the turning points, i.e., the point after which the number of detected bugs remain unchanged for a successive 20 reports.
Note that, the number 20 is set empirically and is just used for demonstrating the trend, not for evaluation purpose. 

The first category is called \textbf{``Rise-Stay''}. We present three example crowdtesting tasks, i.e., P1, P2, and P3 in Figure \ref{fig:observation}.
We can see that for the tasks in this category, with the increase of submitted reports, the percentage of detected bugs would first increase sharply and remain unchanged during the latter part of the task.
For this category, there is an obvious turning point (the red dot in Figure \ref{fig:observation}).
If the crowdtesting platform can close the test task in the turning point, a large portion of cost can be saved without sacrifice the testing quality (i.e., number of detected bugs).
51.8\% (113/218) of our experimental crowdtesting tasks belong to this category.

The second category is called \textbf{``Rise-Stay-Rise''}. The P4, P5, and P6 example tasks in Figure \ref{fig:observation} belong to this category.
We can see that for the tasks in this category, with the increase of submitted reports, the percentage of detected bugs would first increase, and remain unchanged for a noticeable number of reports during the front part of the task, then increase greatly again.
For this category, although the unchanged part of the task is a waste of cost, the task could not be closed at that turning point because there are still a large number of bugs not be reported.
8.7\% (19/218) of our experimental crowdtesting tasks belong to this category.

The third category is called \textbf{``Rise-Stay-Slight Rise''}. We also present three example tasks of this category, i.e., P7, P8, and P9 in Figure \ref{fig:observation}.
We can see that for the tasks of this category, with the increase of submitted reports, the percentage of detected bugs would first increase, remain unchanged in the rear part of the task, and increase slightly.
This category is between the first one and the second one.
Compared with the first category, the tasks of this category does not remain unchanged in the latter part of task.
Compared with the second category, there is only a slight increase in bug number after the turning point.
If the crowdtesting platform close the task in the turning point, the task would be more cost-effective, although a very small portion of bugs would not be detected.
39.5\% (86/218) of our experimental crowdtesting tasks belong to this category.

%We further find that an average of the first 70.5\% (we call it \textit{Knee Point}) submitted reports have detected the total number of bugs mentioned by all reports.
%In addition, for 75\% tasks, the first 62.3\% submitted reports have detected the total number of bugs.
%The reports submitted behind the knee point are almost useless as no new bugs detected, yet consuming nontrivial budgets.
%We suppose the time corresponding to the knee point should be the close time for a crowdsourced testing task.

To summarize, the crowdtesting tasks of the first category and the third category (91.3\% of all experimental tasks) can be closed much earlier than the real-world practice.
This can save the cost of crowdtesting (i.e., less crowd workers are needed), and make it more cost-effective.
Therefore, these findings motivate the necessity and feasibility to conduct the close prediction of crowdtesting tasks. 
% motivates us to automatically predict when to close the crowdtesting task.

\section{Methods for Close Prediction}
\label{sec:approach}

To explore the feasibility of close prediction of crowdtesting task, we adopt the idea from several commonly-used and representative techniques \cite{lyu1996handbook,yamada2014software} and design 8 methods to conduct the close prediction.
% These methods borrow the idea from the most representative and commonly-used techniques \cite{lyu1996handbook,yamada2014software}.

With respect to the context described in Section \ref{subsec:background_crowdtesting}, the designed close prediction method would monitor the report submitting process.
When there is a new report coming, the method would determine whether the submitted reports so far satisfy the predefined stopping criterion; if yes, it determines the time, when the last report was received, as the close point.
The following subsections will introduce each method (including its stopping criterion) in detail. 

% \sout{the close time prediction problem \textit{Ct} is formulated as a dynamic decision problem of 4-tuple \textit{(t, i, R, B)} in our study, where:
%  \textit{t} corresponds to  a crowdtesting task, \textit{i} corresponds to the time interval between the task's open time and its predicted close time, \textit{R} represents the set of submitted reports for task \textit{t} within time interval \textit{i}, and \textit{B} represents the set of unique bugs discovered for task \textit{t} within time interval \textit{i}.} 
 
%  \sout{Suppose that the actual duration for task t is d, the decision is to search for optimal close time for task t, i.e. \textit{Ct}, which satisfies the following objectives:
%  maximizing |B| and minimizing i and minimizing |R|}
%  \junjie{I think it is too strong for this paper}
 
% \sout{This paper will predict the close time in terms of \textit{R}\footnote{Because the predicted close time is in terms of the submitted reports, rather than the physical time, we will use ``close point'' and ``close time'' interchangeably.}, i.e., the stopping criterion is set considering the submitted reports, rather than the physical time; And the other two attributes can be obtained easily after knowing \textit{R}.}

\subsection{Trend Indicator (short for \textit{Trend}) Method}
\label{subsec:method_Trend}

Driven by the flatten-out ending of bug trend discussed in Section \ref{subsec:background_observations}, we design a straightforward and intuitive method for close prediction of crowdtesting tasks.
The basic assumption here is that if the number of accumulated bugs remain unchanged for a while, it may suggest that there is no potential to discover new bugs from that point on. 

More specifically, during the report submitting process, \textit{Trend} method monitors the bugs accumulated so far, and counts the number of consecutive, non-contributing reports (i.e., reports not contributing any newly discovered, unique bugs).
The stopping criterion is that the number of consecutive, non-contributing reports reaches a predefined threshold \textit{stableThres}.

% If at a certain point, the number of non-contributing reports reaches \textit{stableThres} (\textit{stableThres} is a parameter, e.g., 20), \textit{Trend} method predicts the time, when the last report was received, as the close time.

% In detail , when a new report arrives, this method counts
% the number of bugs submitted up to now. If the number of
% bugs remain unchanged for successive stableThres reports
% (stableThres is a parameter), Trend method determines it as
% the close point.

% \sout{when a new report arrives, this method monitors the unique bugs accumulated in submitted reports so far, and counts the number of consecutive, non-contributing reports, meaning that these reports are not contributing any newly discovered, unique bugs. If at certain point, the number of non-contributing reports reaches a \textit{stableThres} value (\textit{stableThres} is a parameter), \textit{Trend} method will recommend that corresponding point of time, i.e. when the last report was received, as the predicted close point for the task under testing.}

\subsection{Peak Indicator (short for \textit{Peak}) Method}
\label{subsec:method_Arrival}

In software reliability researches, Rayleigh's defect arriving model is a classical method to characterize the events of defect detection/arrival following a Rayleigh distribution, i.e. a specialized variant in the Weibull probability distribution family \cite{kan2002metrics}. 
It has been demonstrated effective in predicting the dynamic defect arrival probability w.r.t. specific testing intervals (i.e. days, weeks, months, etc.), as well as the total number of defects associated with a software system based on the cumulative probability distribution \cite{kan2002metrics}.
% It supposes the defects would arrive following the Rayleigh distribution.

\begin{figure}[h!]
\centering
\includegraphics[width=6cm]{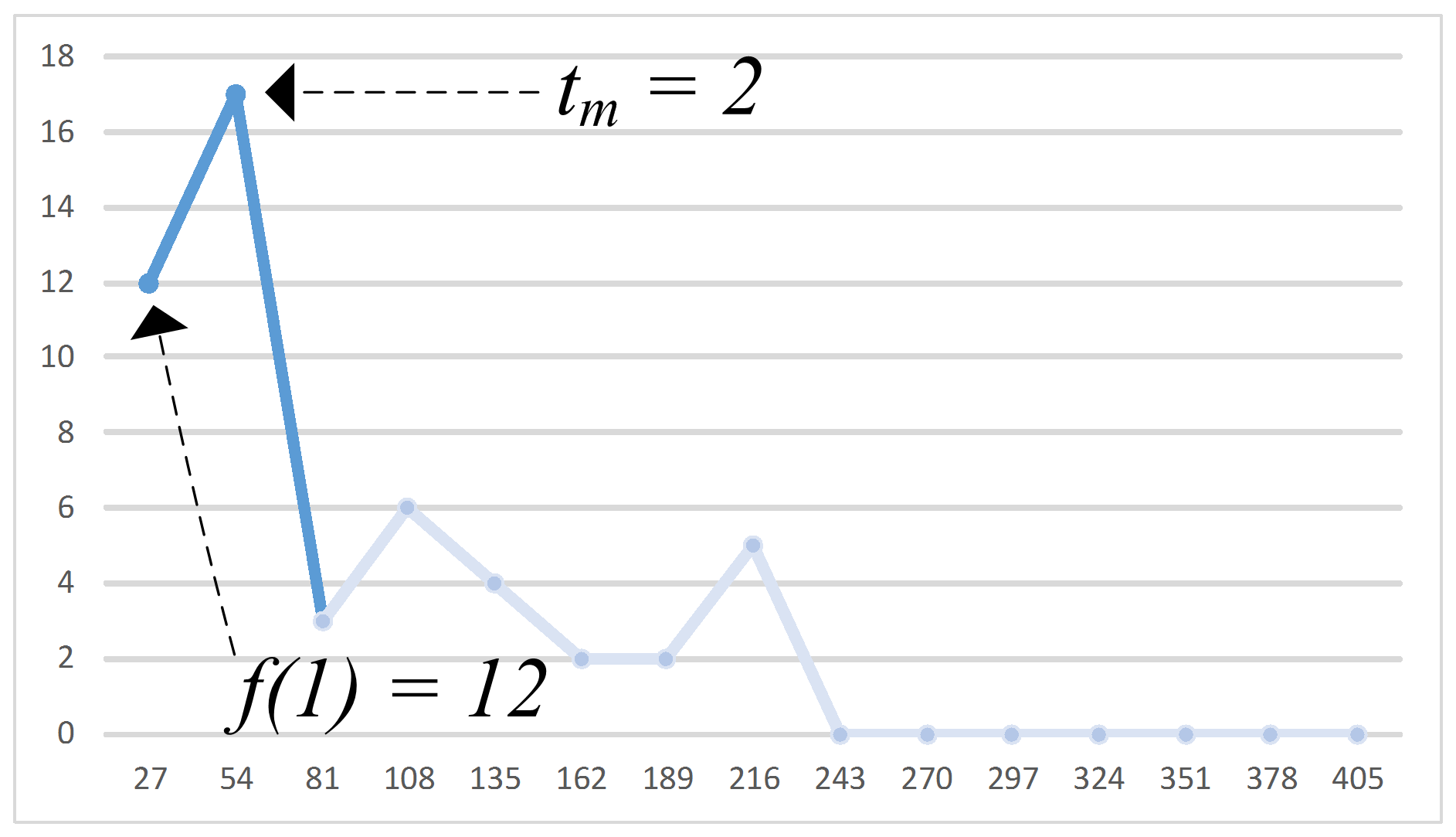}
\caption{Illustrative example of Peak method}
\label{fig:arrival}
% \vspace{-0.1in}
\end{figure}

Inspired by the Rayleigh model, \textit{Peak} method treats every \textit{stepSize} reports as a unit (\textit{stepSize} is a parameter denoting the size of a group of successive reports), and counts the number of bugs in each unit (denoted as $f(t)$).  %newly-detected 
With the arrival of reports, this method monitors $f(t)$ until it first declines.
It then records how many units have passed when $f(t)$ reaches the peak (i.e., the unit before it first declines, denoted as $t_m$).
% reaches the maximum, and records how many \textit{stepSize} reports have arrived (denoted as $t_m$).
Figure \ref{fig:arrival} presents an illustrative example (\textit{stepSize} = 27) for task P1 (Figure \ref{fig:observation}).
Note that, this figure displays the number of detected bugs during each unit (e.g., every 27 reports), while Figure \ref{fig:observation} shows the number of accumulated bugs.
As shown in Figure \ref{fig:arrival},  $f(t)$ starts to decline at the 3rd unit, then $t_m$ is 2. The total number of bugs can then be estimated using Equation \ref{equ_arrival} \cite{kan2002metrics}.

\begin{equation}\label{equ_arrival}
N = C \times e^{(t^{2}/C)} \times f(t)/2t \mbox{, in it } C = 2 t_m^{2}
\end{equation}

In Equation \ref{equ_arrival}, we follow the most common practice and set $t$ as 1, denoting using the 1st unit to fit the distribution.
\textit{Peak} method will continue monitoring the process, and the stopping criterion is that the estimated total number of bugs equals to the number of accumulated bugs so far. 
% up to the last submitted report, it determines that the corresponding time, i.e. when the last report was received, as the predicted close point.

Note that, traditional software testing usually treats the reports within a fixed period (e.g. one day) as a unit. 
However, our initial analysis shows that in crowdtesting, tasks are typically open for shorter periods of time (e.g. 5 days). We did experiments on grouping reports by various time-based unit\footnote{For more details to these experiment results, we will provide online access in the camera-ready version.} and the prediction performance is rather poor compared with the report-based unit treatment, as introduced above.
% , and the crowdworkers conduct tasks in arbitrary time. 

\subsection{Knee Method}
\label{subsec:method_Knee}

\begin{figure}[h!]
\centering
\includegraphics[width=6cm]{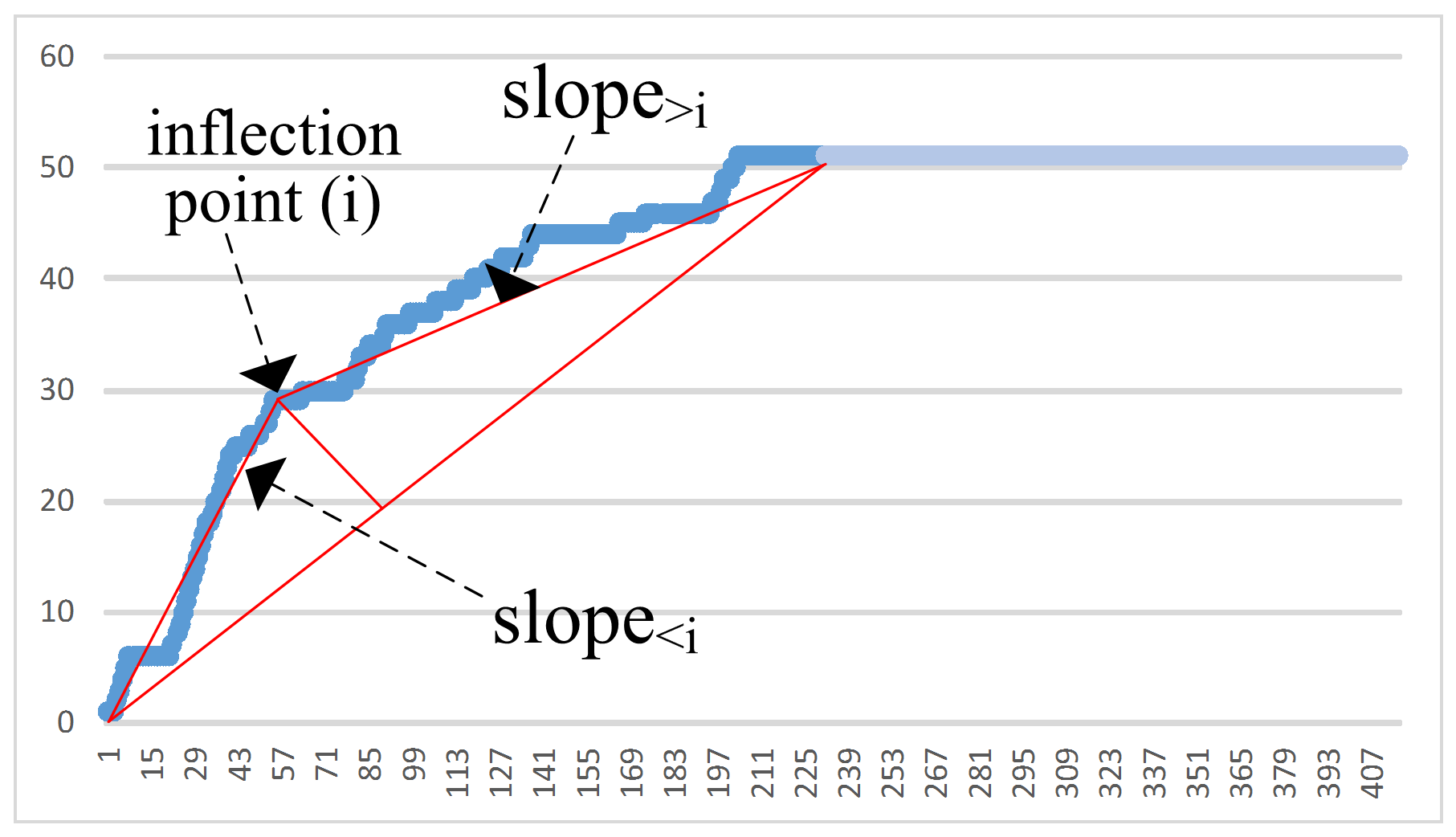}
\caption{Illustrative example of Knee method}
\label{fig:knee}
% \vspace{-0.1in}
\end{figure}

\textit{Knee} method is widely used in technology-assisted review to decide when to stop, considering the quality and reliability \cite{Kneecormack2016engineering}.
This method is based on the the slope of accumulated bug trend curve formed by the reports submitted up to now, as shown in Figure \ref{fig:observation}. It first detects the inflection point $i$ of current curve.
This is done by connecting the starting point and end point of current bug trend curve, then the inflection point is the point which is the farthest from connection line. 
Figure \ref{fig:knee} illustrates an example for task P1 (Figure \ref{fig:observation}) when receiving 232 reports. 
It then compares the slopes before and after $i$, and the stopping criterion is that the ratio of $slope_{<i}/slope_{>i}$ is greater than a specific threshold \textit{kneeThres}.
% \textit{Knee} method determines the time, when the last report was received, as the close point.
% Note that, the bug curve in \textit{Arrival} method is in terms of detected bugs in each unit, while the bug curve in \textit{Knee} method is in terms of the cumulated number of detected bugs.
%For details about knee method, please refer to Cormack and Crossman~\cite{Cormack2016Engineering}.

\subsection{M0 Method}
\label{subsec:method_M0}

We first present some background knowledge shared by the following five methods, i.e., in Section \ref{subsec:method_M0}-\ref{subsec:method_MtCH}.

The Capture-ReCapture (CRC) method, which uses the overlap generated by multiple captures to estimate the toal population, has been applied in software inspections to estimate the total number of bugs \cite{rong2017towards,liu2015adoption,chun2006estimating,mandala2012application}. Existing CRC models can be categorized into four types according to bug detection probability (i.e. identical vs. different) and crowdworker's detection capability (i.e. identical vs. different), as shown in Table \ref{tab:CRC}.

\textit{M0} supposes all different bugs and crowdworkers have the same detection probability. Model \textit{Mh} supposes that the bugs have different probability of being detected. \textit{Mt} supposes that the crowdworkers have different detection capabilities. \textit{Mth} supposes different detection probabilities for different bugs and crowdworkers.

\begin{table}[!h]
\caption{Capture-ReCapture models}
\label{tab:CRC}
\centering
\scalebox{0.95}{
\scriptsize
\begin{tabular}{p{1.95cm}p{1cm}|p{1.8cm}p{2.4cm}}
 &  & \multicolumn{2}{c}{\bfseries{Crowdworker's detection capability}} \\
 & & Identical & Different \\
  \hline
\bfseries{Bug detection} & Identical  & M0 (\textit{Sec\ref{subsec:method_M0} M0})  & Mt (\textit{Sec\ref{subsec:method_MtCH} MtCH})  \\
\bfseries{probability}   & Different & Mh (\textit{Sec\ref{subsec:method_MhJK} MhJK, Sec\ref{subsec:method_MhCH} MhCH})  & Mth (\textit{Sec\ref{subsec:method_Mth} Mth}) \\
\end{tabular}
}
%\multirow{2}{*}{
%\vspace{-0.05in}
\end{table}

Based on the four basic CRC models, various estimators were developed. % to implement the calculations.
According to a recent systematic review \cite{liu2015adoption}, \textit{MhJK}, \textit{MhCH}, \textit{MtCH} are the three most frequently investigated and most effective estimators in software engineering.
Apart from that, we investigate another two estimators (i.e., \textit{M0} and \textit{Mth}) to ensure all four basic models are investigated.
Base on the general idea of these five models, we design five corresponding methods to be applied to the close prediction problem for crowdtesting tasks in this section and Section \ref{subsec:method_Mth}-\ref{subsec:method_MtCH}.

Method \textit{M0} treats every \textit{capSize} reports as a capture (\textit{capSize} is a parameter denoting how many reports are considered in each capture).
At the end of each capture (i.e., the number of received reports is the multiple of \textit{capSize}), it conducts the following operations.
It treats the newest capture as the second round, while all previous captures as the first round.
It then counts the number of bugs in the first round (denoted as $n_1$) and number of bugs in the second round (denoted as $n_2$), as well as the number of bugs contained in both rounds (i.e., duplicate reports between 1st round and 2nd round, denoted as $m$).
Note that, when counting $n_2$, for the bug duplicated with the one in 1st round, we still treat it as a bug because the 2nd round is considered as a recapture. 
The total number of bugs is estimated as Equation \ref{equ_CRC} \cite{M0laplace1783naissances}.
The stopping criterion is that the predicted total number of bugs is equal with the actual number of detected bugs so far.
Note that, we simply treat $n_1 + n_2$ as the total number when $m$ is 0.
% , \textit{M0} method determines the time, when the last report was received, as the close point.

% \vspace{-0.1in}
\begin{equation}\label{equ_CRC}
N = \frac{n_1 \times n_2}{m}
% , \mbox{ } ( N = n_1 + n_2, \mbox{ if } m=0)
\end{equation}

\subsection{Mth Method}
\label{subsec:method_Mth}

This method treats every \textit{capSize} reports as a capture.
%  (i.e., reports number is the multiple of \textit{capSize})
At the end of each capture, \textit{Mth} method estimates the total number of bugs based on Equation \ref{equ_mth}, \ref{equ_mth3} \cite{Mthlee1996estimating}.
The stopping criterion is the same with \textit{M0} method, i.e., the predicted total number of bugs is equal with the actual number of detected bugs so far.
% , \textit{Mth} method determines it as the close point.

\begin{equation}\label{equ_mth}
\footnotesize{ N = \frac{D}{C} + \frac{f_1}{C}\gamma^2 } \mbox{, } \footnotesize{ C = 1 - \frac{f_1}{\sum_{k=1}^{t}kf_k} } \\
\end{equation}
%\vspace{-0.05in}

\begin{equation}\label{equ_mth3}
\footnotesize{ \gamma^2 = max\{{\frac{\frac{D}{C}\sum_{k}k(k-1)f_k}{2\sum\sum_{j<k}n_jn_k}-1, 0}\} }
\end{equation}

In it, $N$ is the predicted total number of bugs;

$D$ is the actual number of bugs captured so far;

$t$ is the number of capture;

$n_j$ is the number of bugs detected in each capture; Note that, as in \textit{M0}, we do not consider the bug duplication among different captures.

$f_k$ is the number of bugs captured exactly $k$ times in all captures, i.e., $\sum f_i = D$.

\textbf{Note that, these value assignments are shared among the following methods.}

\subsection{MhJK Method}
\label{subsec:method_MhJK}

\textit{MhJK} method is similar with \textit{Mth} method, except its equation for estimating the total number of bugs in Equation \ref{equ_mhjk} \cite{MhJKburnham1978estimation}.

\begin{equation}\label{equ_mhjk}
\footnotesize{ N = D + \frac{t-1}{t}f_1 }
\end{equation}

Note that, the \textit{MhJK} estimation has three other expressions. 
We use all four expressions, and choose the right estimator through hypothesis testing as suggested in \cite{MhJKburnham1978estimation}.
Please refer to \cite{MhJKburnham1978estimation} for more details.

\subsection{MhCH Method}
\label{subsec:method_MhCH}

\textit{MhCH} method is similar with \textit{Mth} method, except its equation for estimating the total number of bugs in Equation \ref{equ_mhch}, \ref{equ_mhch2} \cite{MhCHchao1988estimating}.

\begin{equation}\label{equ_mhch}
\footnotesize{ N = D + \frac{f_1^2}{2f_2} }
\end{equation}

or
\begin{equation}\label{equ_mhch2}
\footnotesize{ N = D + \frac{[\frac{f_1^2}{2f_2}][1-\frac{2f_2}{tf_1}]}{1 - \frac{3f_3}{tf_2}}, \mbox{ if } tf_1 > 2f_2, tf_2 > 3f_3, 3f_1f_2 > 2f_2^2 }
\end{equation}

\subsection{MtCH Method}
\label{subsec:method_MtCH}

\textit{MtCH} method is also similar with \textit{Mth} method, except its equation for estimating the total number of bugs in Equation \ref{equ_mtch} \cite{MtCHchao1987estimating}.

\begin{equation}\label{equ_mtch}
\footnotesize{ N = D + \frac{\sum_{i=1}^{t} \sum_{j=i+1}^{t}Z_iZ_j}{f_2+1} }
\end{equation}

In it, $Z_i$ is the number of bugs detected only in the $i_{th}$ capture, i.e., $\sum Z_i = f_1$.

\section{Experiment Design}
\label{sec:experiment}

To evaluate the effectiveness of the proposed close prediction methods, we design a series of evaluation experiments. This section presents the research questions, evaluation metrics, and setup for the experiments.

\subsection{Research Questions}
\label{subsec:design_rq}

We formulate two research questions to be addressed in the experiment:
% Our evaluation addresses the following two research questions:

\begin{itemize}
\item \textbf{RQ1}: How does the parameter of each method influence the prediction performance, and what is the optimal parameter value?
\end{itemize}
%\vspace{-0.05in}

Since each method has a predefined parameter, RQ1 aims at analyzing the sensitivity of the  parameters on prediction performance, and investigating its feasibility to converge on the optimal parameter value which optimizing the prediction performance. 

\begin{itemize}
\item \textbf{RQ2}: How effective is each method in close prediction of crowdtesting task?
% How effective are the proposed methods in order to improve decision making on close prediction of crowdtesting task?
\end{itemize}
%\vspace{-0.05in}

RQ2 aims at evaluating the performance of the proposed methods to prove their effectiveness in improving current crowdtesting practices in determining optimal task closing.

% \begin{itemize}
% \item \textbf{RQ3}: How stable is each method in close prediction of crowdtesting task?
% \end{itemize}

% Results of RQ1 demonstrate that the parameter can influence the performance, RQ3 is to evaluate the stability of each method under pre-tuned parameters.

Note that, because there is no existing work for close prediction of crowdtesting, we do not have explicit baselines.
The only possible baseline, i.e., the actual practice indicated by the total number of detected bugs and the total number of submitted reports, is reflected in the evaluation metrics since \textit{\%bug} and \textit{\%reducedCost} are compared with actual practice.

\subsection{Evaluation Metric}
\label{subsec:design_metric}

% Given a crowdtesting task, after the close prediction we can obtain the tuple \textit{(D, R, B)} about the duration, submitted reports (i.e., cost paid), and detected bugs (i.e., outcomes).
% The evaluation focuses on \textit{R} and \textit{B} because we find that the cost paid and testing outcomes are more valuable than the time duration, through talking with the project managers in our experimental platform. 
We measure the performance of each close prediction method based on how much percentage of bugs can be detected together with how much percentage of cost can be saved.

\textbf{\%bug} is the percentage of bugs detected by the predicted close point.
We treat the number of historical detected bugs as the total number.
% Note that, ``bug'' here are referred as no duplicate bugs.
The larger \textit{\%bug}, the more bugs can be detected by the predicted clost point, the more effective the corresponding close prediction method is.

\textbf{\%reducedCost} is the percentage of saved cost by the predicted close point.
To derive this metric, we first obtain the percentage of reports submitted at the close point, in which we treat the number of historical submitted reports as the total number.
We suppose this is the percentage of consumed cost\footnote{The most important cost in crowdsourced testing is the reward for workers, and their submitted reports are usually equally paid \cite{cui2017multi,cui2017who}; Hence we suppose it is reasonable for using the number of submitted reports to indicate the consumed cost.} and \%reducedCost is derived using 1 minus the percentage of consumed cost.
% the ratio of saved cost (i.e. the difference between total budgeted cost and consumed cost) and the total budgeted cost. 
The larger \textit{\%reducedCost}, the more cost is saved and less testing cost is needed, the more effective the proposed method is.

Intuitively, an increase in \%bug would be accompanied with a decrease in \%reducedCost.
Motivated by the F1 (or F-Measure) in prediction approaches of software engineering \cite{nam2017heterogeneous,wang2017domain,wang2016towards}, we further define \textbf{F1}, to measure the harmonic mean of \%bug and \%reducedCost as follows: 
\begin{equation}
\footnotesize{ F1 = \frac{2 \times \%bug \times \%reducedCost}{\%bug + \%reducedCost} }
\end{equation}

% $(2 \times \%bug \times \%reducedCost) / (\%bug + \%reducedCost) $), providing a single performance measure.

\begin{figure*}[t!]
  \centering
  \begin{subfigure}{0.24\textwidth}
    \includegraphics[width=4.8cm]{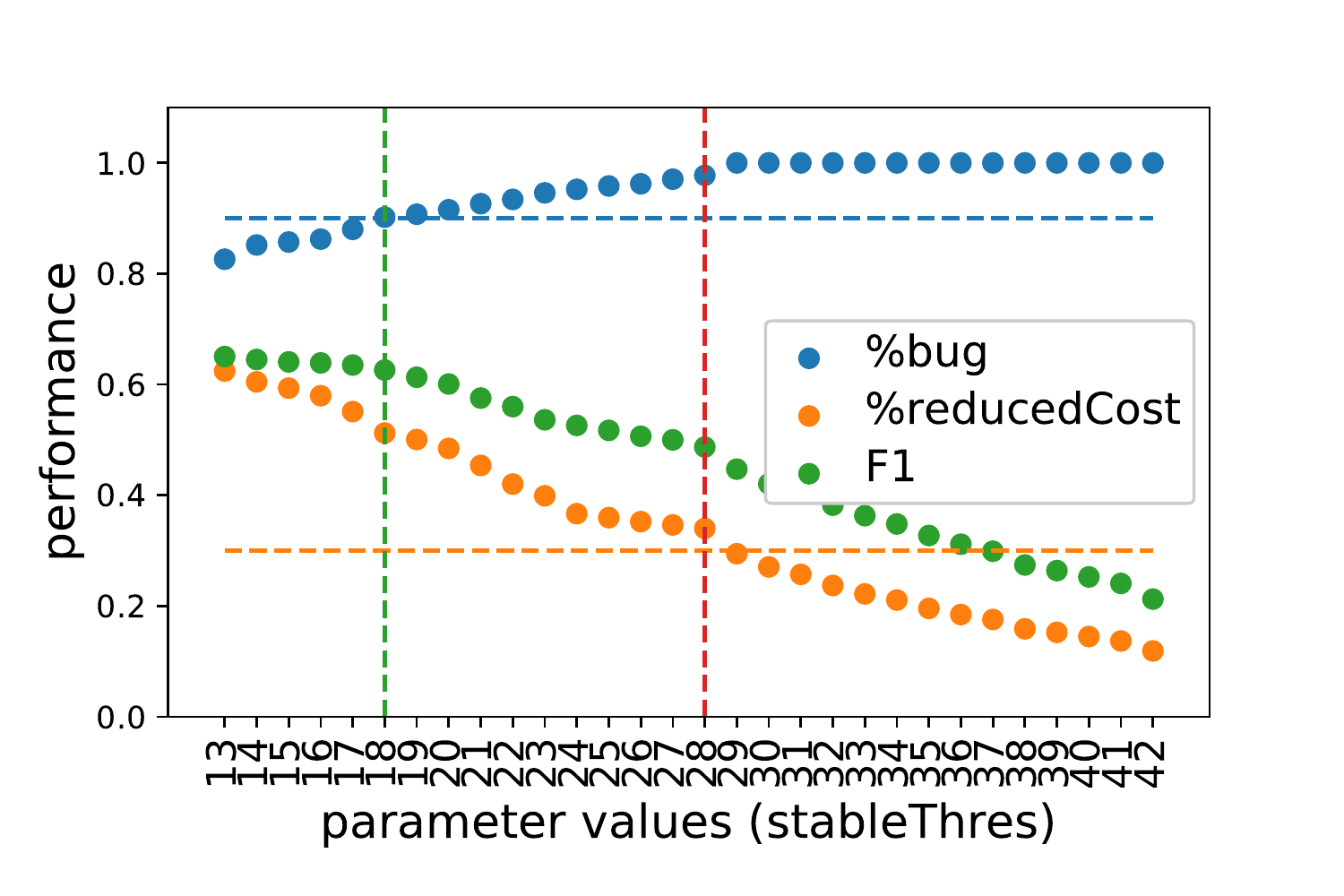}
	 \caption{Trend}
	 \label{fig:TLTrend}
  \end{subfigure}
  \begin{subfigure}{0.24\textwidth}
    \includegraphics[width=4.8cm]{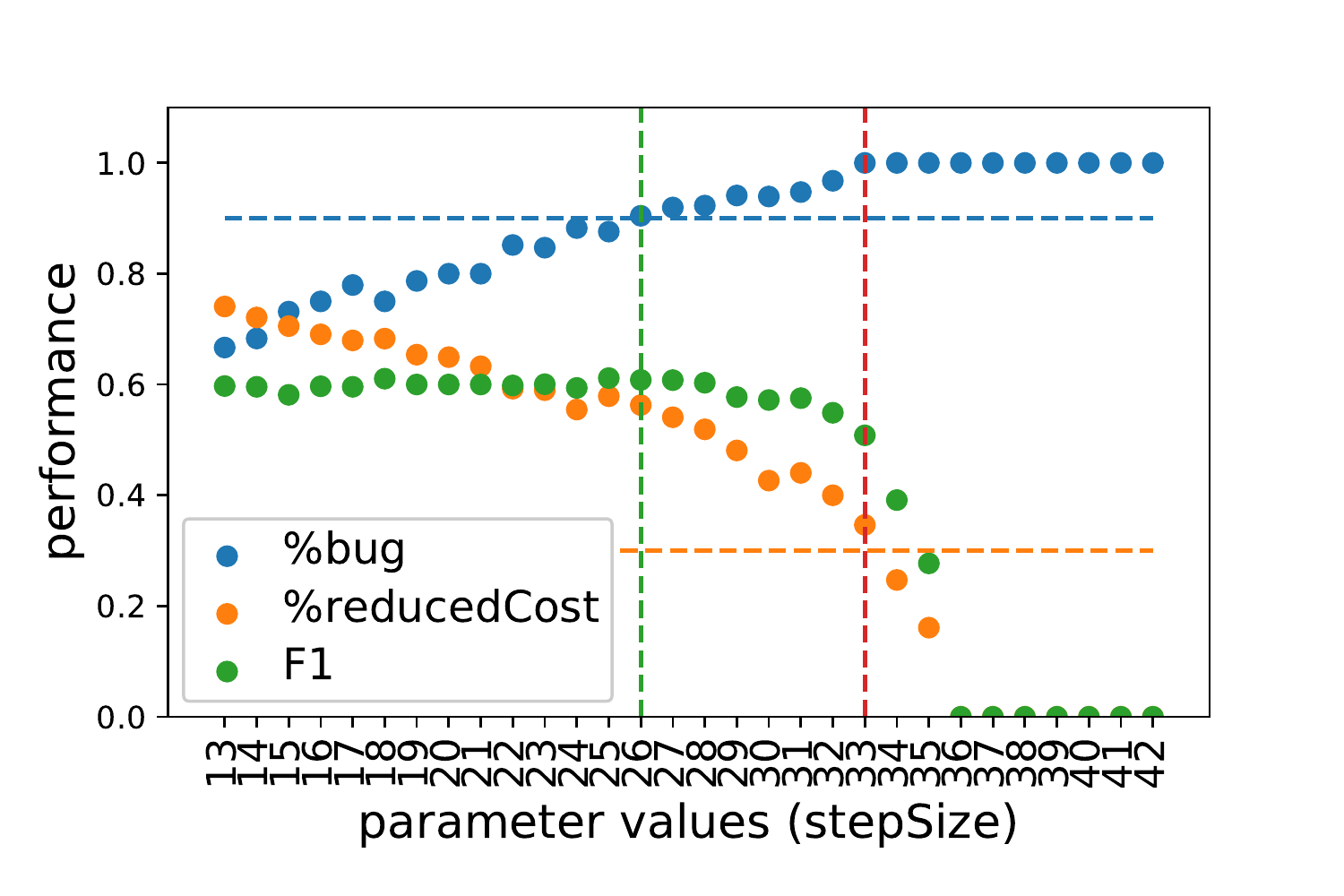}
	\caption{Peak}
   \label{fig:TLArrival}
  \end{subfigure}
  \begin{subfigure}{0.24\textwidth}
    \includegraphics[width=4.8cm]{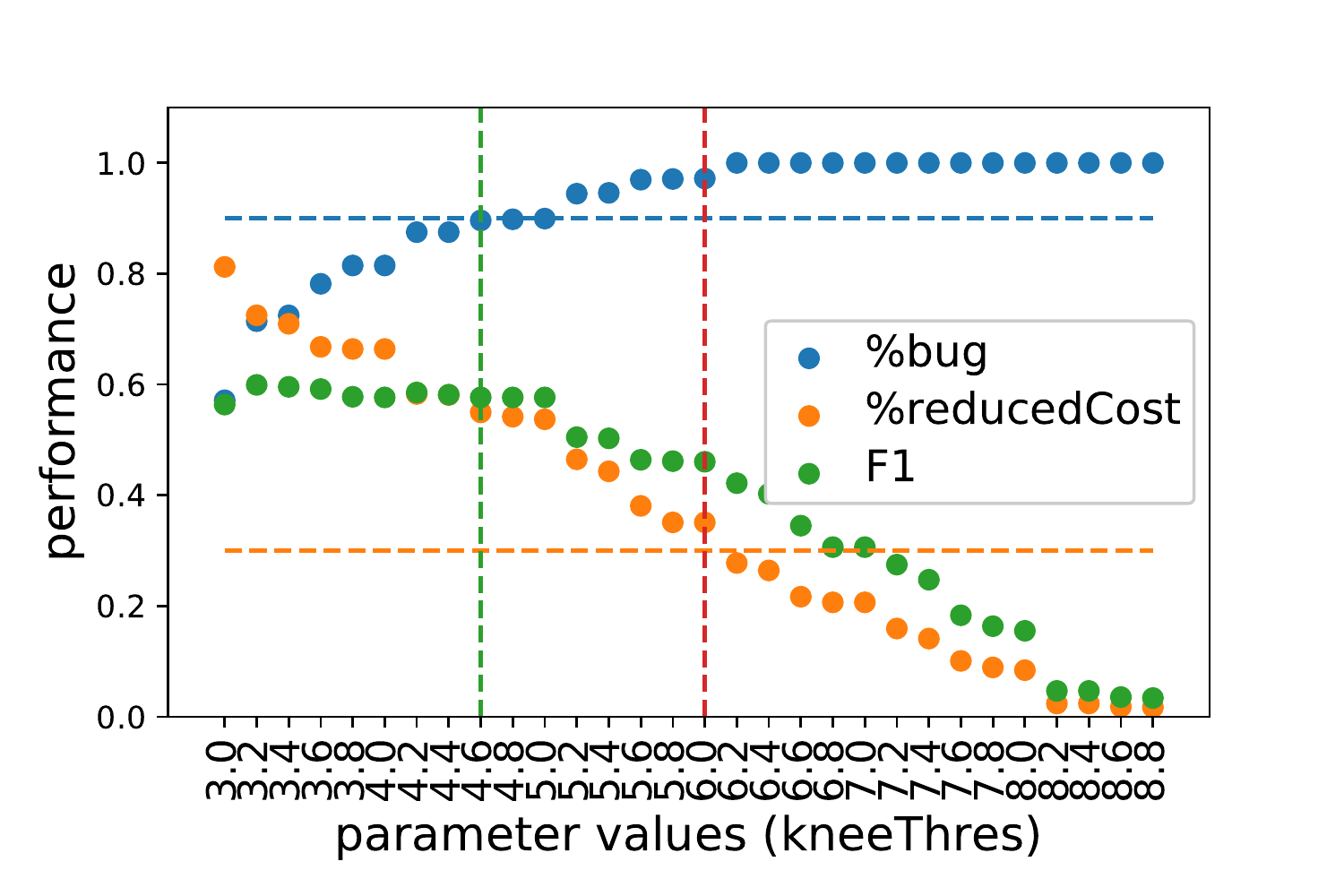}
	\caption{Knee}
    \label{fig:TLKnee}
  \end{subfigure}
   \begin{subfigure}{0.24\textwidth}
    \includegraphics[width=4.8cm]{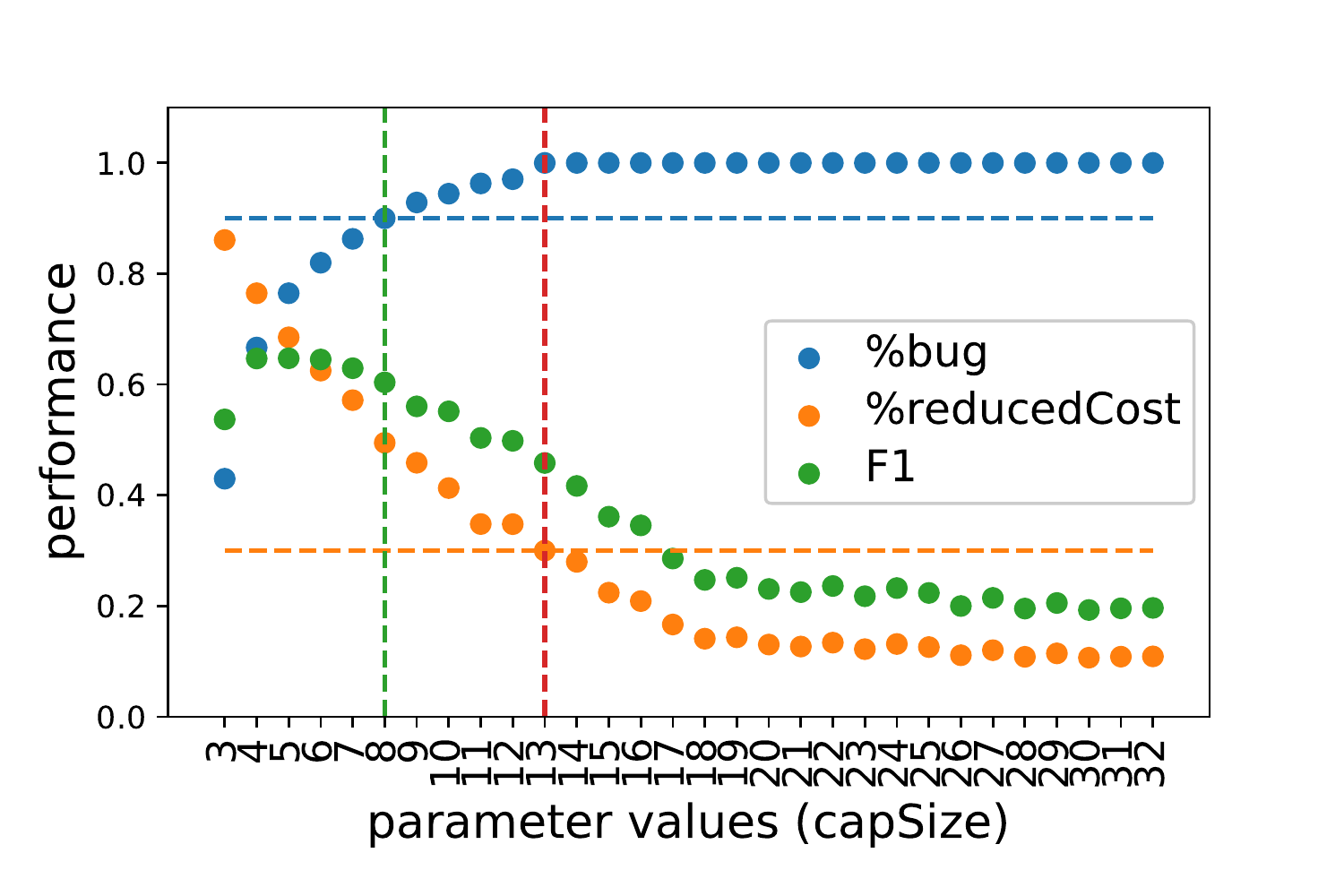}
	\caption{M0}
    \label{fig:TLM0}
  \end{subfigure}
   \begin{subfigure}{0.24\textwidth}
    \includegraphics[width=4.8cm]{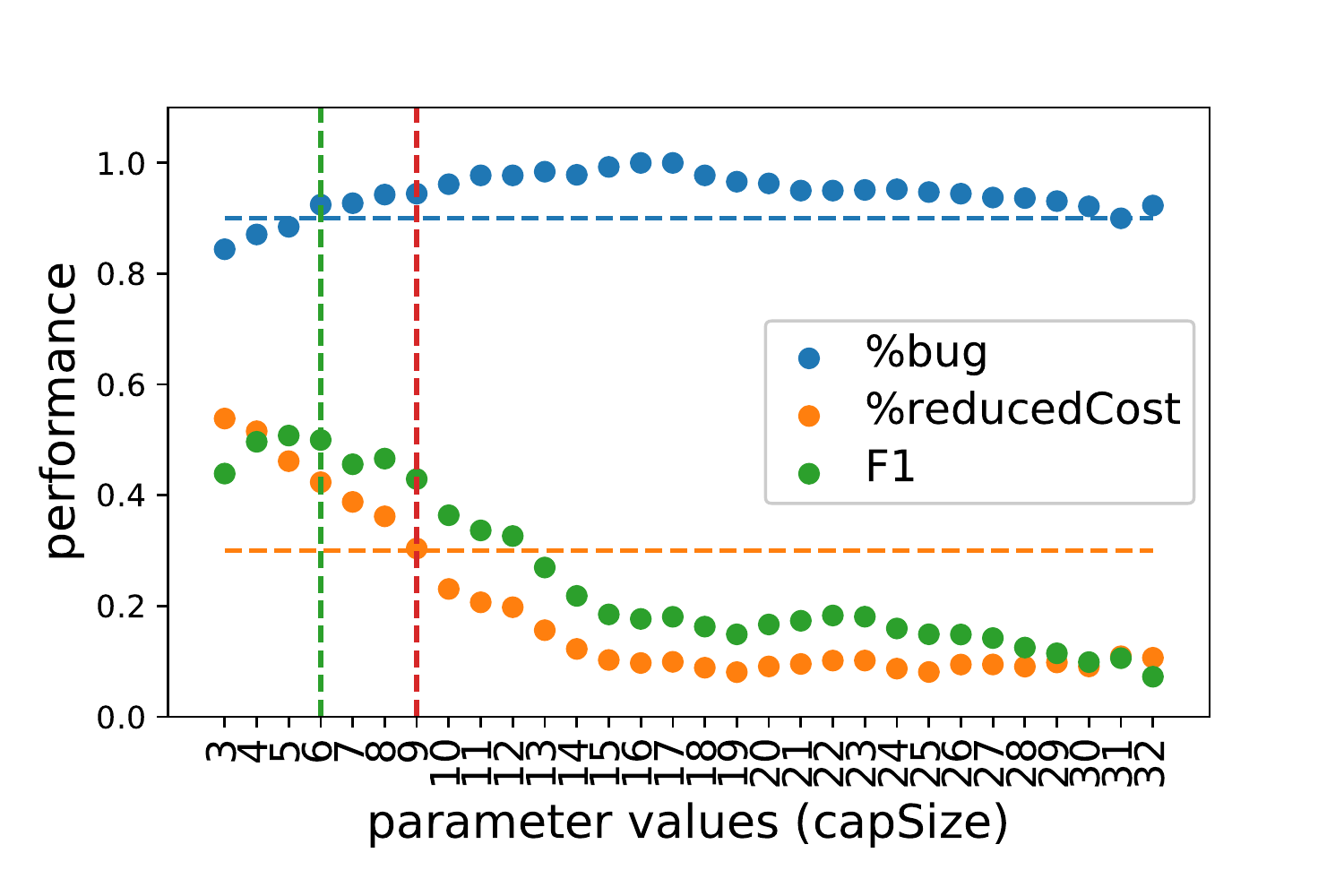}
	 \caption{Mth}
	 \label{fig:TLMth}
  \end{subfigure}
  \begin{subfigure}{0.24\textwidth}
    \includegraphics[width=4.8cm]{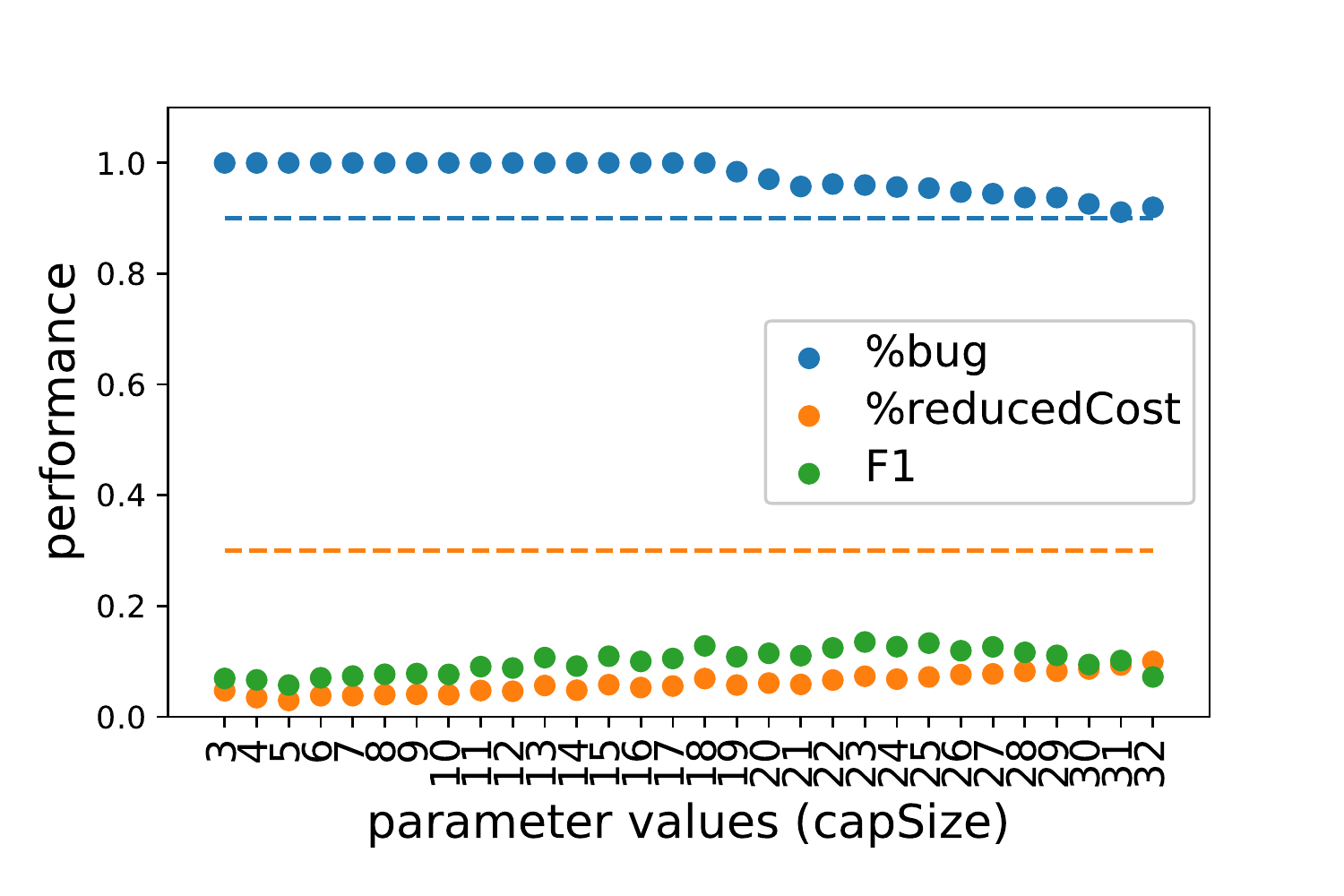}
	\caption{MhJK}
   \label{fig:TLMhJK}
  \end{subfigure}
  \begin{subfigure}{0.24\textwidth}
    \includegraphics[width=4.8cm]{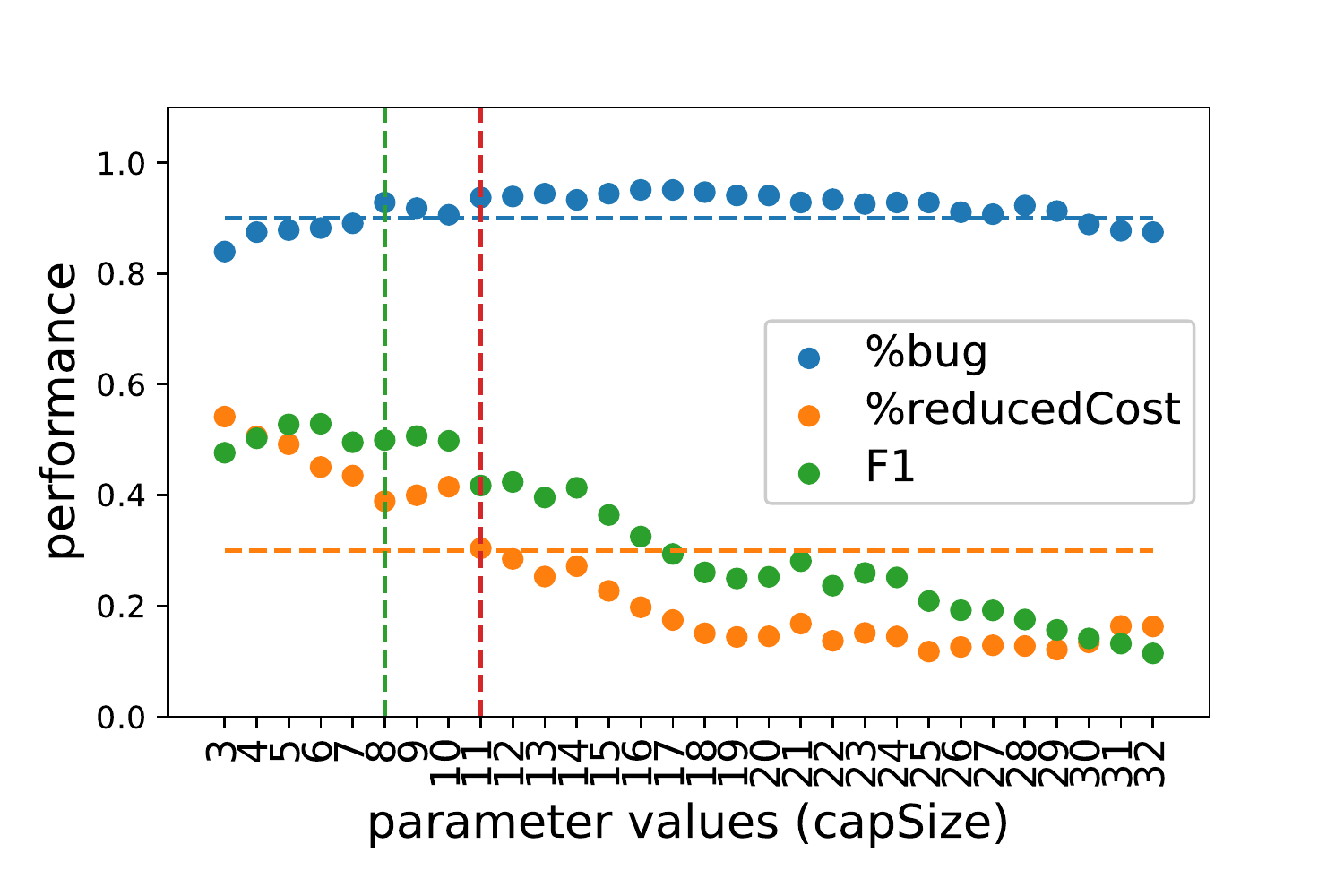}
	\caption{MhCH}
    \label{fig:TLMhCh}
  \end{subfigure}
   \begin{subfigure}{0.24\textwidth}
    \includegraphics[width=4.8cm]{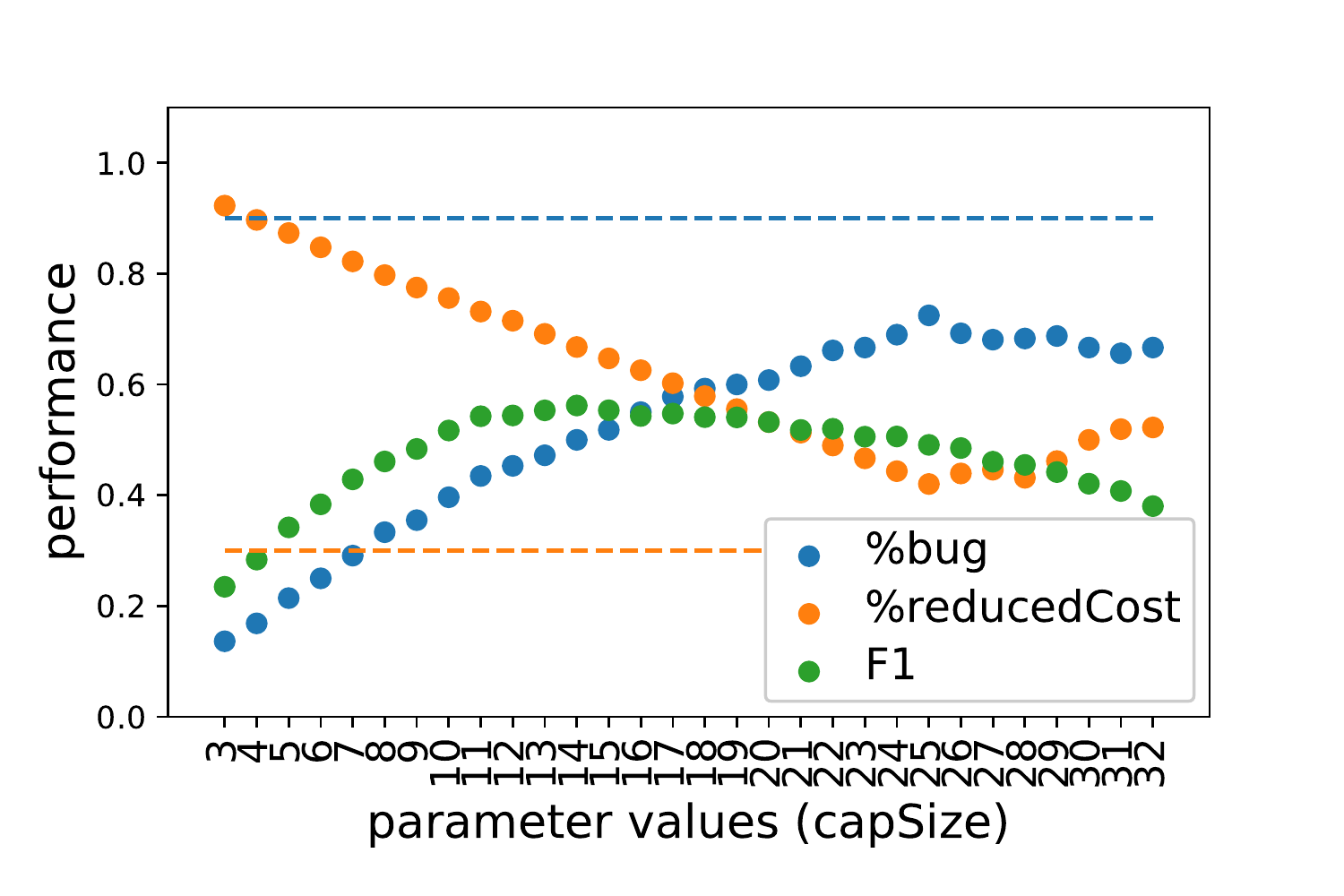}
	\caption{MtCH}
    \label{fig:TLMtCh}
  \end{subfigure}
  \caption{Influence of parameter values on prediction performance, i.e., sensitivity analysis (RQ1)}
  \label{fig:thres}
% \vspace{-0.1in}
\end{figure*}

\begin{figure}[t!]
\centering
\includegraphics[width=8.9cm]{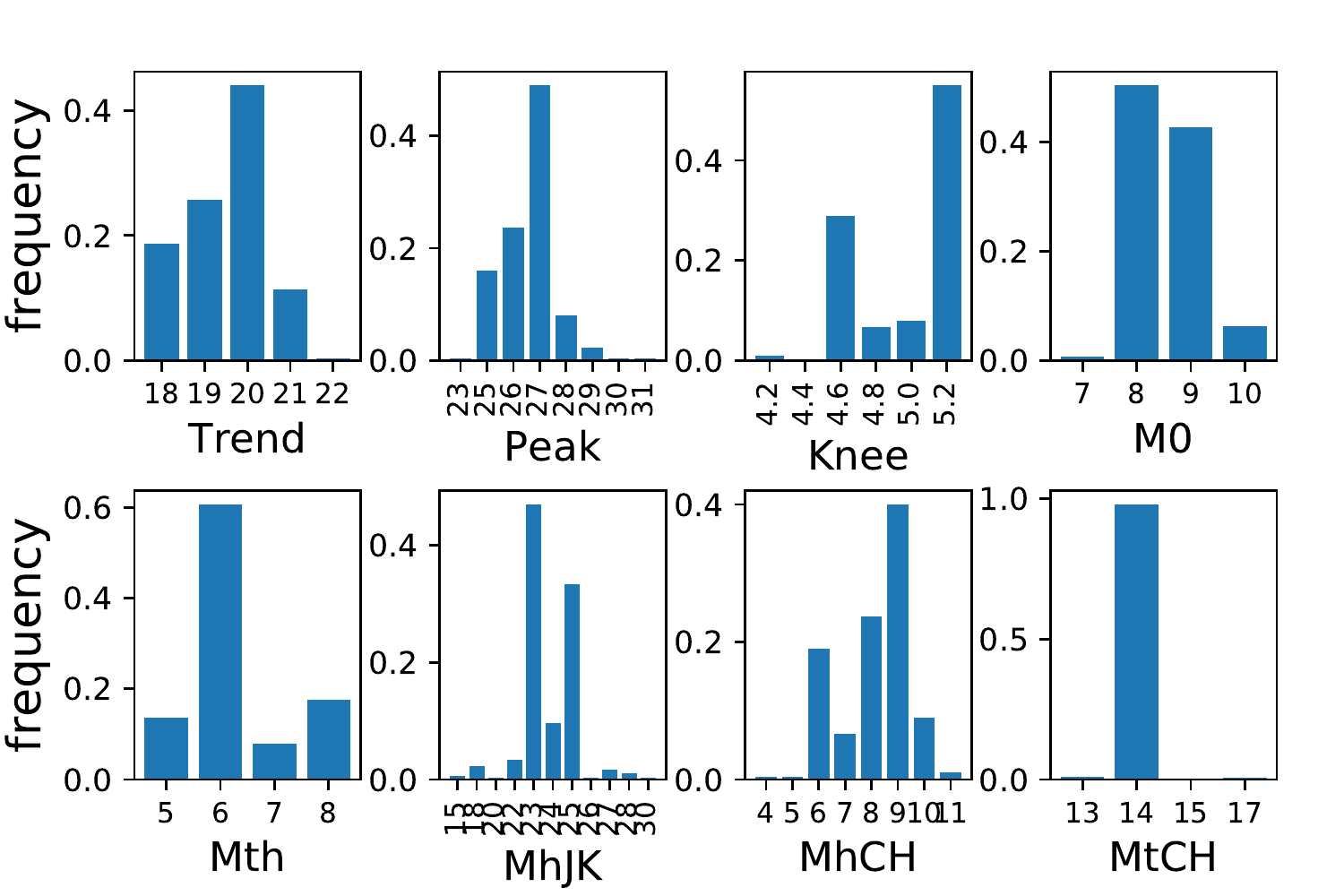}
\caption{Tuned optimal parameter values (RQ1)}
\label{fig:parameter}
%\vspace{-0.1in}
\end{figure}

\subsection{Experimental Setup}
\label{subsec:design_setup}

For RQ1, to demonstrate the influence of parameter and determine the optimal parameter value for each method, we first experiment with all candidate parameter values for each method on all experimental crowdtesting tasks,
and then obtain the median performance across all tasks under each candidate parameter value (results are shown in Section \ref{subsubsec:RQ1-1}).
Next, we conduct three-fold cross validation \cite{data_ming} and repeat 1000 times to alleviate the randomness.
In each cross validation, we randomly separate the {\task} experimental crowdtesting tasks into three equal folds.
We employ each two folds as training set to tune the optimal parameter value (rules for parameter tuning will be shown in Section \ref{subsubsec:RQ1-2}), and use the remaining one fold as testing set. 
We then record the tuned optimal parameter value, and obtain the frequency  of these values appeared in the 1000 cross validations, which will be reported in Section \ref{subsubsec:RQ1-3}.  

% The tuned optimal parameter values will be ranked their frequencies in the 1000 cross validations, which will be reported in Section \ref{subsubsec:RQ1-3}. 
% \junjie{We donot rank}

For RQ2, we use the parameter value which appeared most frequently (i.e., the highest bar in Fig. \ref{fig:parameter}) in the 1000 cross validations to investigate the effectiveness of each method.

\section{Results and Analysis}
\label{sec:result}
This section presents the results and analysis from the evaluation experiments.

\subsection{Answering RQ1: Parameter Sensitivity and Tuning of Optimal Parameter Values}
\label{subsec:RQ1}

\subsubsection{\textbf{Parameter Sensitivity Analysis}}
\label{subsubsec:RQ1-1}

Figure \ref{fig:thres} shows the scatter plots between parameter values and prediction performance for the 8 close prediction methods.
% , where in each chart, the x-axis represents different parameter value settings in the experiment, and y-axis displays the corresponding mean performance in 1000 runs of the corresponding method.
Note that, for each method, we explored 100 candidate parameter values, and the charts present 30 of them with relatively good performance. The rest are omitted in this paper due to space limit.

In each chart of Fig. \ref{fig:thres}, three dotted curves demonstrate the prediction performance, i.e. \textit{\%bug}, \textit{\%reducedCost}, and \textit{F1}.
Generally speaking, with the increase of  \textit{\%bug},  \textit{\%reducedCost} would decrease.
This is consistent with common intuition that it costs more to detect every additional bugs.

It's also observed that, almost for all methods, the change in performance is quite smooth.
Put it in another way, the performances only demonstrate small difference with two continuous parameter values.
This indicates that choosing two adjacent parameter values would not bring much variation in the performance.
This is valuable when applying these methods in real-world practice, which will be discussed in details in Section \ref{subsubsec:RQ1-3}.
% We will show why this is a valuable result in Section \ref{subsec:RQ3}.

Furthermore, for each method, only under certain parameter values, the prediction can achieve a satisfying performance.
 %(i.e., \textit{\%bug} larger than 90\% and \textit{\%reducedCost larger than 30\%}).
For example, for \textit{Peak}, if the parameter value is smaller than 21, the percentage of detected bugs is not so high, i.e., less than 0.8 (Figure \ref{fig:TLArrival}).
For \textit{Knee}, the reduced cost is quite low, i.e., less than 0.05, when the parameter value is larger than 8.2 (Figure \ref{fig:TLKnee}).
%the parameter value should range from 20 to 28 (Figure \ref{fig:TLTrend}), while the parameter value should range from 26 to 33 (Figure \ref{fig:TLArrival}).
%For method \textit{MhJK} and \textit{MtCH}, the prediction performance is barely satisfactory under all parameter values.
This implies that there is a need to tune the optimal parameter value when using these methods in real-world practices.

\subsubsection{\textbf{Rules for Determining Optimal Parameter Value}}
\label{subsubsec:RQ1-2}

Presumably, the optimal parameter value will not only lead to a sufficiently good F1, but also ensure a pair of satisfactory values for \textit{\%bug} and  \textit{\%reducedCost} which conveys meaningful and actionable insights to be easily interpreted and applied in planning for crowdtesting practices.
It is observed from Figure \ref{fig:thres} that under certain parameter value, although F1 is the highest, only 68\% bugs can be detected with saving 76\% cost (Figure \ref{fig:TLM0}, when parameter value is 5).
We do not think this prediction makes much sense in real-world crowdtesting context, because a large portion of bugs still remain undetected.
Similarly, we also hope the  \textit{\%reducedCost} should satisfy a lower-bound restriction to ensure the achievement of cost-effectiveness objective.
% for potential task requesters looking into adopting crowdtesting.
%under some parameter values, although F1 is the highest and \%bugs can achieve 100\%, the saved effort is only 20\% (Figure \ref{fig:TLMth}, ).
%We also argue the practicability of such kind of prediction.

In this study, after consulting with the managers from the {\company}, we employ a rule-based approach to help determine the optimal parameter value. 
More specifically, as shown in Fig. \ref{fig:thres}, two horizontal lines are introduced to specify the performance expectation corresponding to the following two rules: \textbf{R1)} a minimum acceptable \textit{\%bug} value of 90\%;  \textbf{R2)} a minimum acceptable \textit{\%reducedCost} value of 30\%.
In other words, following these two rules, we expect to determine the optimal parameter value which will guarantee the prediction performance with a \textit{\%bug} value higher than 90\% and a \textit{\%reducedCost} value greater than 30\%.

Based on these two rules, the determined range of parameter values is depicted using two vertical lines in Fig. \ref{fig:thres}.
The left line (i.e. Green) is identified by applying Rule \textit{R1} to the \textit{\%bug} data, and the right line (i.e. Red) is identified by applying Rule \textit{R2} to the \textit{\%reducedCost} data. 
Finally, we introduce a third rule: \textbf{R3)} the optimal parameter value is the specific parameter value associated with a maximum F1 from the restricted range identified by R1 and R2.
For example, the optimal parameter value of \textit{Peak} is 27 (Figure \ref{fig:TLArrival}), and the optimal parameter value of \textit{M0} is 8 (Figure \ref{fig:TLM0}).

% and 3) choose the optimal parameter value which maximizes the F1-score.
% In another word, following these two rules, we expect to determine the optimal parameter which will guarantee the prediction performance with a \textit{\%bug} value higher than 90\% and a  \textit{\%reducedCost} value greater than 30\%.

% , where the line to the Left (i.e. Blue) line is identified by applying Rule 1 to the \textit{\%bug} data, and the line to the Right (i.e. Red) is identified by applying Rule to to the  \textit{\%reducedCost} data. Finally, we consider the parameter value associated with a maximum F1-score from the restricted range identified by Rule 1 and Rule 2 as the optimal parameter value, for each method. 

Please note that the restriction values (i.e., \textit{90\%, 30\%}) in this experiment is decided based on our discussion with the project managers in {\company}.
People can customize their own restriction values when deciding the optimal parameter value.
For example, if they want to save more cost, they can lower the \%bug restriction to 80\%.
Another note is that, for those methods which do not have parameter value satisfying the restriction (i.e., \textit{MhJK} and \textit{MtCH}), we simply choose the parameter value with the largest F1 (directly applying R3).
We will explain the reason for the low performance of \textit{MhJK} and \textit{MtCH} in Section \ref{subsec:RQ2}.

% Finally, 2 of the 8 methods (i.e., \textit{MhJK} and \textit{MtCH}) do not have parameter value satisfying the above restriction rules. In such cases, we apply Rule R3 directly and choose the parameter value with the largest F1. 

\subsubsection{\textbf{Tuned Parameter Values}}
\label{subsubsec:RQ1-3}

Figure \ref{fig:parameter} demonstrates the frequency of tuned optimal parameter values in the 1000 cross validations (see Section \ref{subsec:design_setup}).
It is noticeable that for most methods, there are mainly 2-4 values determined as the optimal parameter values in different training dataset.
For example, the optimal parameter values for \textit{Trend} is 18 to 21, while the optimal parameter values for \textit{M0} is 8 and 9.
% the tuned optimal parameter values are universal, i.e., 

These tuned optimal parameter values are consistent with the optimal parameter values determined on all experimental tasks in Figure \ref{fig:thres}.
For example, the optimal parameter value (Figure \ref{fig:thres}) of \textit{Peak} and \textit{M0} are respectively 27 and 8, while most frequently tuned optimal parameter value (Figure \ref{fig:parameter}) of \textit{Peak} and \textit{M0} are also 27 and 8.

In addition, these optimal parameter values are adjacent with each other, for example, 18, 19, 20, 21 for \textit{Trend}.
We have mentioned in Section \ref{subsubsec:RQ1-1}, the performances obtained by adjacent parameter values usually have small difference.
This indicates that based on the high frequency parameter values in Figure \ref{fig:parameter}, the performance might not exert much difference among the different choices (e.g. using 19 or 20 in \textit{Trend}).

This experimental finding provides insightful guidelines for applying these methods.
Crowdtesting managers can use the high frequency parameter values obtained in our experimental dataset, and different choice might not bring large variance in the performance. Of course, if they have historical crowdtesting tasks, they can apply additional tuning of the optimal parameter value.

% For all the designed methods, our 1000 cross experiments show that the optimal parameter falls on a sharp range. it means that these high frequency parameter is universal and could be used commonly. % re-tuning the optimal parameter is also recommended.

In the following subsection, for each method, we will report results from experiments applying the most frequent parameter value to investigate the effectiveness of the method.

\subsection{Answering RQ2: Effectiveness}
\label{subsec:RQ2}

\begin{figure}[t!]
\centering
\includegraphics[width=9cm]{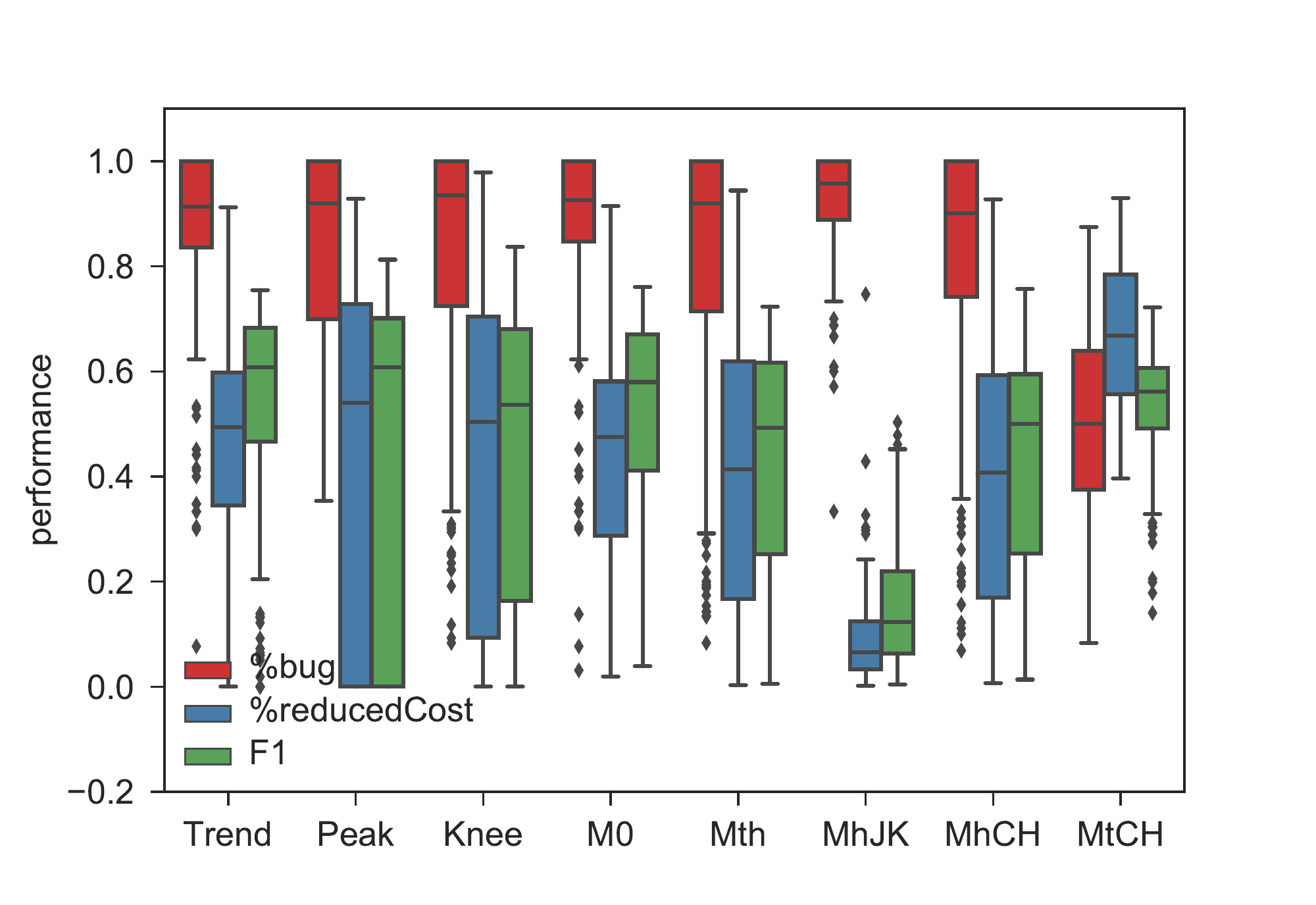}
\caption{Performance on 218 crowdtesting tasks (RQ2)}
\label{fig:effective}
% \vspace{-0.1in}
\end{figure}

\begin{table}[!tb]
\scriptsize
\caption{Statistical performance for effectiveness (RQ2)}
\label{tab:mean-effective}
\centering
\begin{tabular}{p{1.2cm}|p{0.8cm}p{0.8cm}p{0.8cm}|p{0.8cm}p{0.8cm}p{0.8cm}}
\hline
& \multicolumn{3}{c|}{Median} & \multicolumn{3}{c}{Std.} \\
\hline
 &  \%B & \%R & F1 & \%B & \%R & F1  \\
\hline
Trend & 0.913 & 0.494 & 0.607 & 0.173 & 0.202 &	0.181 \\
Peak & 0.919 & 0.540 & 0.607 & 0.180 & 0.345 & 0.333\\
Knee & 0.934 & 0.503 & 0.536 & 0.236 &	0.317 &	0.274 \\
M0 & 0.926 & 0.475 & 0.579 & 0.187 & 0.212 & 0.186\\
Mth & 0.919 & 0.414 & 0.493 & 0.251 & 0.269 & 0.220\\
MhJK & 0.956 & 0.065 & 0.123 & 0.099 &	0.084 &	0.106\\
MhCH & 0.900 & 0.407 & 0.499 & 0.246 & 0.257 &	0.210\\
MtCH & 0.500 & 0.667 & 0.561 & 0.177 & 0.130 &	0.105\\
\hline
\end{tabular}
\vspace{-0.1in}
\end{table}

\begin{table*}[!th]
\scriptsize
\caption{Results of Mann-Whitney U Test for effectiveness (RQ2)}
\label{tab:test-effective}
\centering
% \scalebox{0.9}{
\begin{tabular}{p{0.7cm}|p{0.4cm}|p{0.4cm}|p{0.4cm}|p{0.45cm}|p{0.45cm}|p{0.45cm}|p{0.4cm}|p{0.4cm}|p{0.4cm}|p{0.45cm}|p{0.45cm}|p{0.45cm}|p{0.45cm}
|p{0.45cm}|p{0.45cm}|p{0.45cm}|p{0.45cm}|p{0.45cm}|p{0.45cm}|p{0.45cm}|p{0.45cm}}
\hline
 & \multicolumn{3}{c|}{Peak} & \multicolumn{3}{c|}{Knee} & \multicolumn{3}{c|}{M0}
  & \multicolumn{3}{c|}{Mth} & \multicolumn{3}{c|}{MhJK} & \multicolumn{3}{c|}{MhCH} & \multicolumn{3}{c}{MtCH} \\
\hline
& \%B & \%R & F1 & \%B & \%R & F1 & \%B & \%R & F1 & \%B & \%R & F1 & \%B & \%R & F1 & \%B & \%R & F1 & \%B & \%R & F1 \\
\hline
Trend & 0.97 & 0.69 & 0.07 & 0.70 & 0.79 & 0.00\tiny{*} & 0.33 & 0.20 & 0.08 & 0.59 & 0.01\tiny{*} & 0.00\tiny{*} & 0.00\tiny{*} & 0.00\tiny{*} & 0.00\tiny{*} & 0.25 & 0.00\tiny{*} & 0.00\tiny{*} & 0.00\tiny{*} & 0.00\tiny{*} & 0.00\tiny{*} \\
\hline
Peak & & &  & 0.92 & 0.16 & 0.81 & 0.51 & 0.83 & 0.30 & 0.38 & 0.46 & 0.17 & 0.00\tiny{*} & 0.00\tiny{*} & 0.00\tiny{*} & 0.14 & 0.67 & 0.12 & 0.00\tiny{*} & 0.00\tiny{*} & 0.21 \\
\hline
Knee & & &  & & &  & 0.74 & 0.75 & 0.06 & 0.43 & 0.36 & 0.06 & 0.01\tiny{*} & 0.00\tiny{*} & 0.00\tiny{*} & 0.13 & 0.26 & 0.06 & 0.00\tiny{*} & 0.00\tiny{*} & 0.44 \\
\hline
M0 & & &  & & &  & & &  & 0.24 & 0.13 & 0.00\tiny{*} & 0.00\tiny{*} & 0.00\tiny{*} & 0.00\tiny{*} & 0.06 & 0.06 & 0.00\tiny{*} & 0.00\tiny{*} & 0.00\tiny{*} & 0.09 \\
\hline
Mth & & &  & & &  & & &  & & &  & 0.00\tiny{*} & 0.00\tiny{*} & 0.00\tiny{*} & 0.50 & 0.93 & 0.98 & 0.00\tiny{*} & 0.00\tiny{*} & 0.00\tiny{*} \\
\hline
MhJK & & &  & & &  & & &  & & &  & & &  & 0.00\tiny{*} & 0.00\tiny{*} & 0.00\tiny{*} & 0.00\tiny{*} & 0.00\tiny{*} & 0.00\tiny{*} \\
\hline
MhCH & & &  & & &  & & &  & & &  & & &  & & &  & 0.00\tiny{*} & 0.00\tiny{*} & 0.00\tiny{*} \\
\hline
%MtCH & & &  & & &  & & &  & & &  & & &  & & &  & & &  \\
%\hline
\end{tabular}
% }
\footnotesize {\protect\newline Note that we mark the values less than 0.05 with \textbf{*}, denoting the difference is significant.}
% \vspace{-0.1in}
%\vspace{-0.1in}
\end{table*}

% We conduct cross validation and repeat 100 times to avoid randomness.
% We employ the specific cross validation with the median F1 among the 100 validations to investigate the effectiveness of each method.
Figure \ref{fig:effective} demonstrates the prediction performance on 218 experimental crowdtesting tasks with each method.
Table \ref{tab:mean-effective} additionally demonstrates the mean and standard deviation of \textit{\%bug}, \textit{\%reducedCost} and \textit{F1} for better illustration.
In addition, Table \ref{tab:test-effective} presents the p-value of Mann-Whitney U Test between each two methods.

At first glance, we can see that the first four methods (i.e., \textit{Trend}, \textit{Peak}, \textit{Knee}, and \textit{M0}) achieve relatively better performance, while the performance of the last four methods are a little worse.
This is beyond our expectation, because three of the last four methods (i.e., \textit{MhJK, MhCH, MtCH}) have proven to be the best bug estimators in software inspection researches \cite{liu2015adoption,briand2000comprehensive,chun2006estimating,rong2017towards}.
In crowdtesting, these capture-recapture methods are worse than the simple capture-recapture methods (i.e., M0).
This might because in software inspection, the inspectors are predetermined and they test under closed environment.
For crowdtesting, there is no predefined workers for a task, and each registered crowdworker of the platform can come at any time, so they test under open environment.
This implies the well-designed capture-recapture algorithms (i.e., \textit{MhJK}, \textit{MhCH}, and \textit{MtCH}) for closed environment might not be suitable for crowdtesting.

For the four better methods (i.e., \textit{Trend}, \textit{Peak}, \textit{Knee}, and \textit{M0}), we first put our focus on the median performance on the experimental dataset.
Generally speaking, the performance of them do not exert significant difference (all the p-value, except between \textit{Trend's F1} and \textit{Knee's F1}, is larger than 0.05 in Table \ref{tab:test-effective}).
This implies that all these four methods can obtain a similar median performance on our experimental crowdtesting tasks.

We then shift our focus on the variance of the performance.
From Figure \ref{fig:effective} and Table \ref{tab:mean-effective}, we can easily see that the standard deviation of \textit{Peak} and \textit{Knee} is much larger than the variance of \textit{Trend} and \textit{M0}.
This implies for a noticeable portion of our experimental crowdtesting tasks, the performance obtained by \textit{Peak} and \textit{Knee} is low.
Put it in another way, on our experimental crowdtesting tasks, the performance obtained by \textit{Trend} and \textit{M0} are more stable, thus more effective. 
% Note that, ``stable'' here is in terms of different experimental tasks, while ``stable'' in next section is in terms of different parameter values.

To summarize, 1)\textit{Trend} and \textit{M0} are the best two methods for close prediction in terms of median and deviation performance across all experimental tasks; 2) \textit{Trend} is slightly better than \textit{M0}, considering median \textit{F1} of \textit{Trend} is 0.607, while median \textit{F1} of \textit{M0} is 0.579.
This is out of our expectation, because \textit{Trend} is the most straightforward and intuitive method.
This might because the bug detection process of crowdtesting task is more open and complicated than traditional software testing process.
Under a complicated scenario, a simple way can usually take effect \cite{brighton2006robust,fu2017easy}.
This suggests, to conduct the close prediction of crowdtesting task, one should first try the most simple method.

Furthermore, under the best method \textit{Trend}, a median of 91\% bugs can be detected with 49\% reduced cost.
This implies the task requester can save 49\% of budgeted cost with the risk of missing 9\% bugs. 
The reduced cost is a tremendous figure when considering the large number of tasks delivered in a crowdtesting platform.

\section{Discussion}
\label{sec:discussion}

\subsection{Rise-Stay-Rise Pattern Causes Performance Bottleneck}
\label{subsec:dis_badcase}

We conduct additional analysis on the potential causes for the tasks with low \textit{\%bug}.
% large errors in the predicted close point, as compared with the actual close point. 
The initial results suggest that for all methods, the low prediction performance is attributed to the errors from the crowdtesting tasks belonging to the second category, i.e. whose reports arrival follows the Rise-Stay-Rise pattern (see Section \ref{subsec:background_observations}). Intuitively, during the ``Stay'' phase, the number of bugs remain unchanged for a noticeable large number of consecutive reports, which consequently misleads the close prediction methods to determine it as the close point. All 8 proposed methods suffer from this problem.

One possible mitigation is to combine these methods with other estimators which are based on orthogonal assumption other than the leverage of dynamic defect arrival pattern (i.e., methods proposed in this paper). 
For example, it might be helpful to predict the total number of bugs based on the function point or other related features using machine learner \cite{fu2017easy,nam2017heterogeneous,menzies2007data}. Then this number can be used as a sanity check to detect false alarm when our designed method predicts a close point with much smaller total number of bugs. 
% it as close, it is more likely to be a misleading flat area.
We will explore more techniques to detect and address this problem in future work.

\begin{figure*}[t!]
  \centering
% \hspace{-0.2in}
  \begin{subfigure}{0.24\textwidth}
    \includegraphics[width=4.8cm]{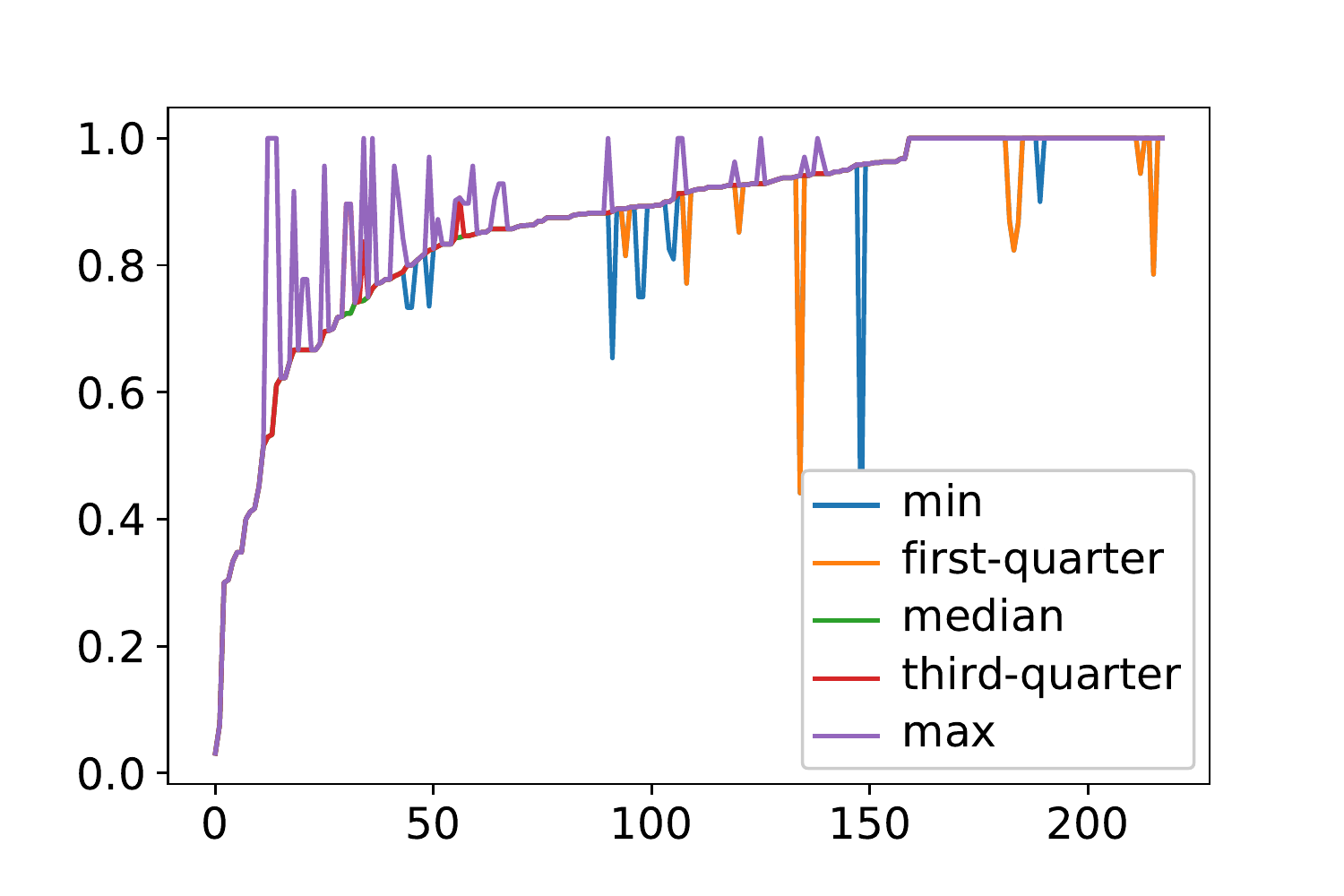}
	 \caption{Trend \%bug}
	 \label{fig:trend-b}
  \end{subfigure}
%  \hspace{-0.1in}
  \begin{subfigure}{0.24\textwidth}
    \includegraphics[width=4.8cm]{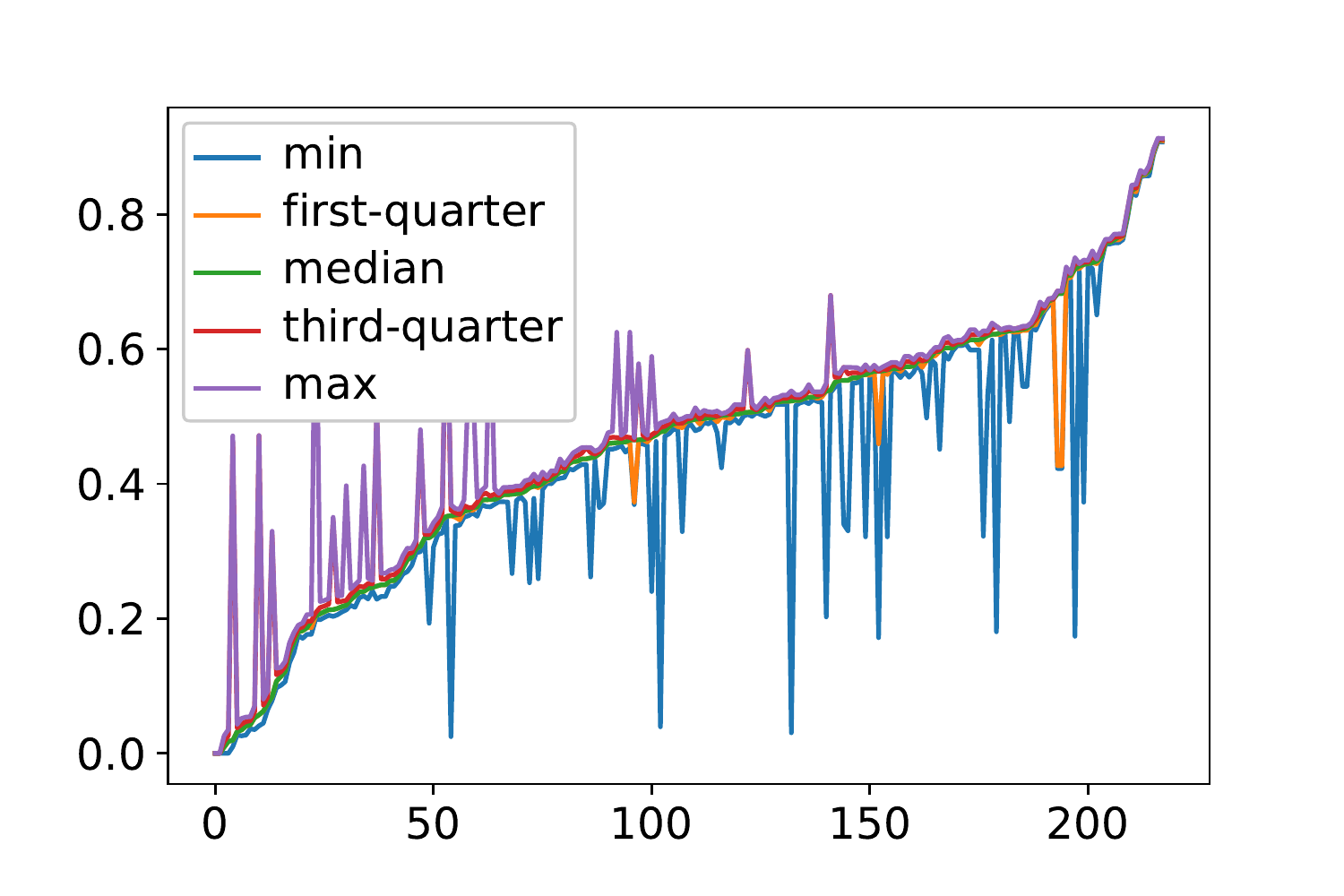}
	\caption{Trend \%reducedCost}
   \label{fig:trend-c}
  \end{subfigure}
  \begin{subfigure}{0.24\textwidth}
    \includegraphics[width=4.8cm]{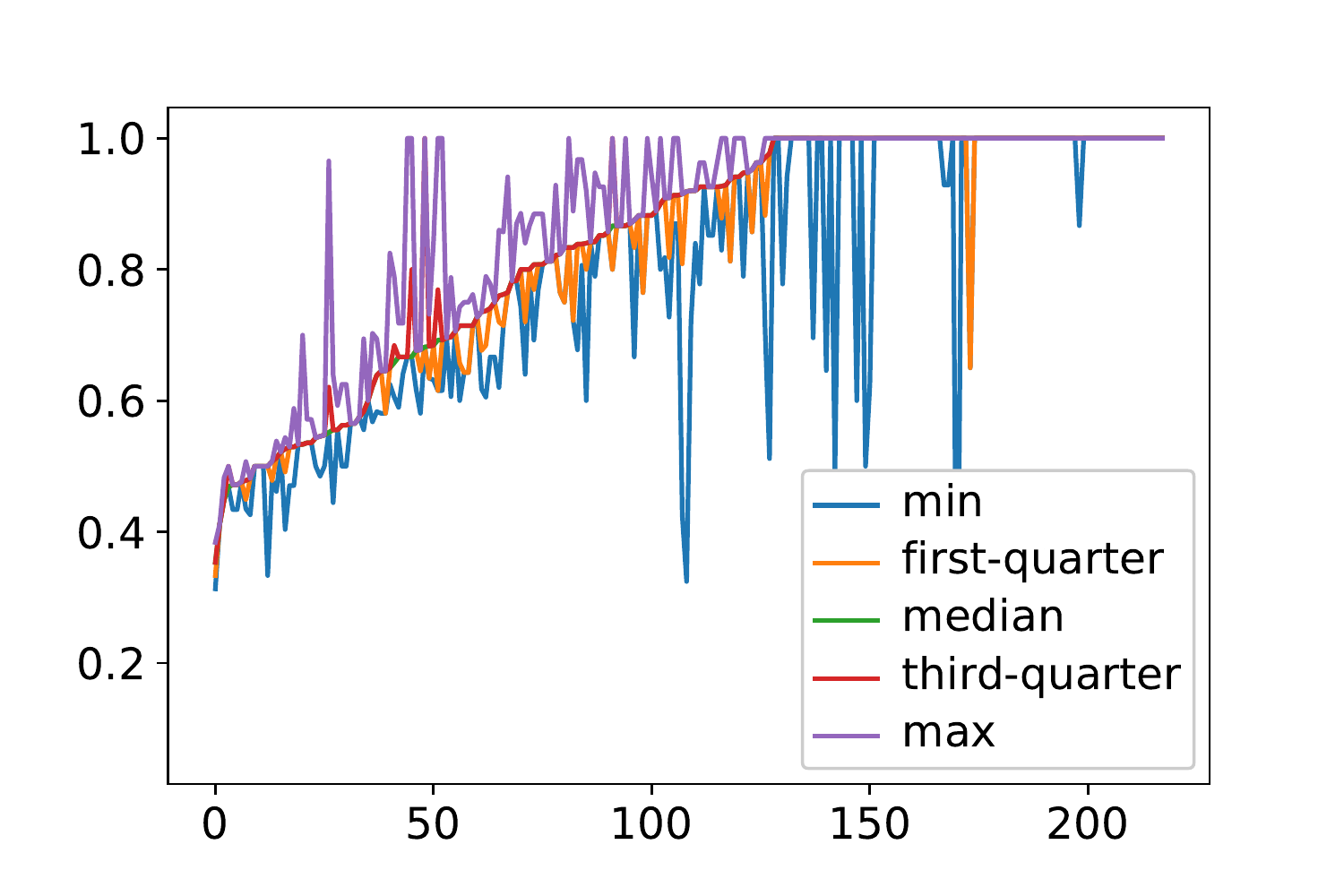}
	\caption{Peak \%bug}
    \label{fig:peak-b}
  \end{subfigure}
  \begin{subfigure}{0.24\textwidth}
    \includegraphics[width=4.8cm]{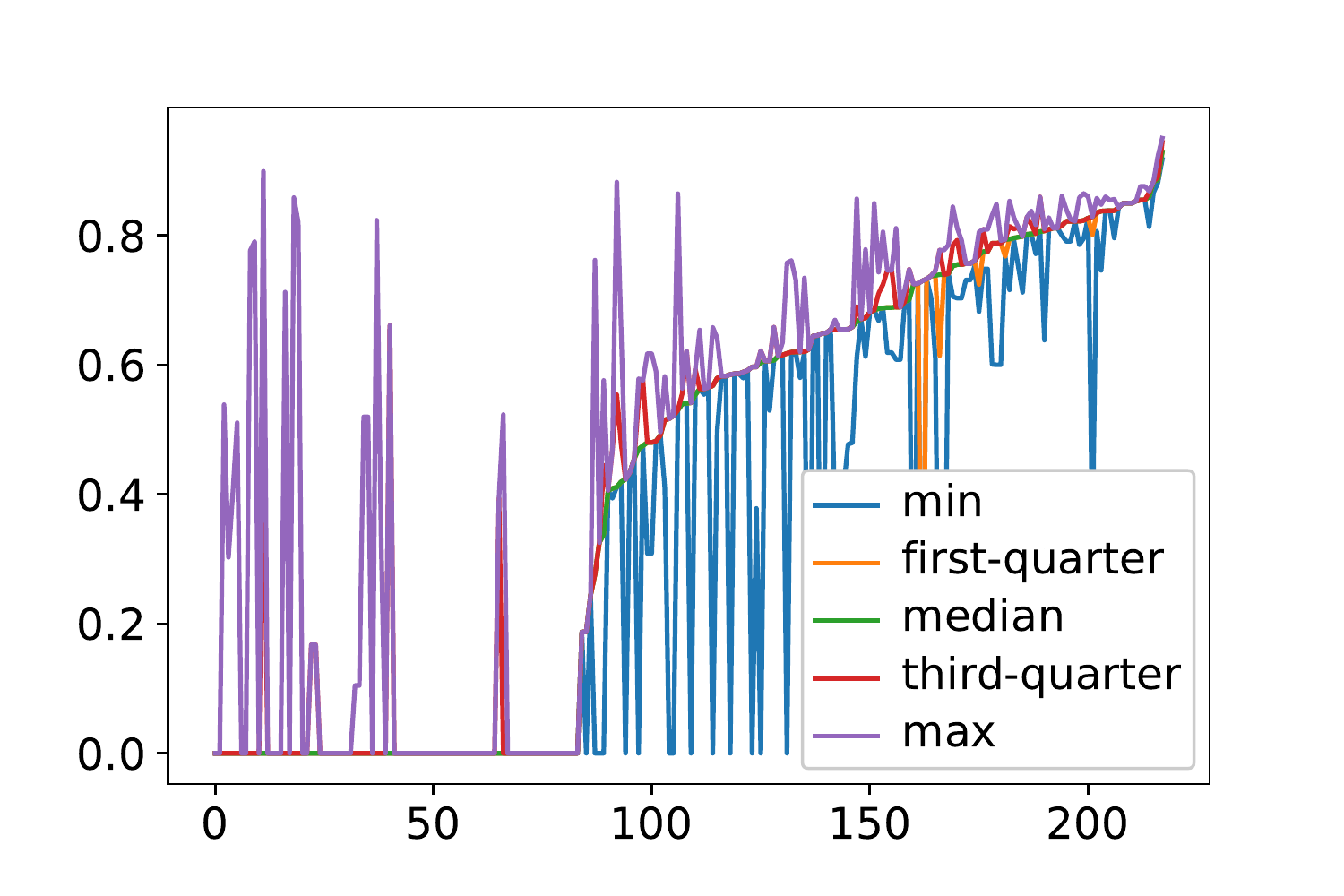}
	\caption{Peak \%reducedCost}
    \label{fig:peak-c}
  \end{subfigure}
  \begin{subfigure}{0.24\textwidth}
    \includegraphics[width=4.8cm]{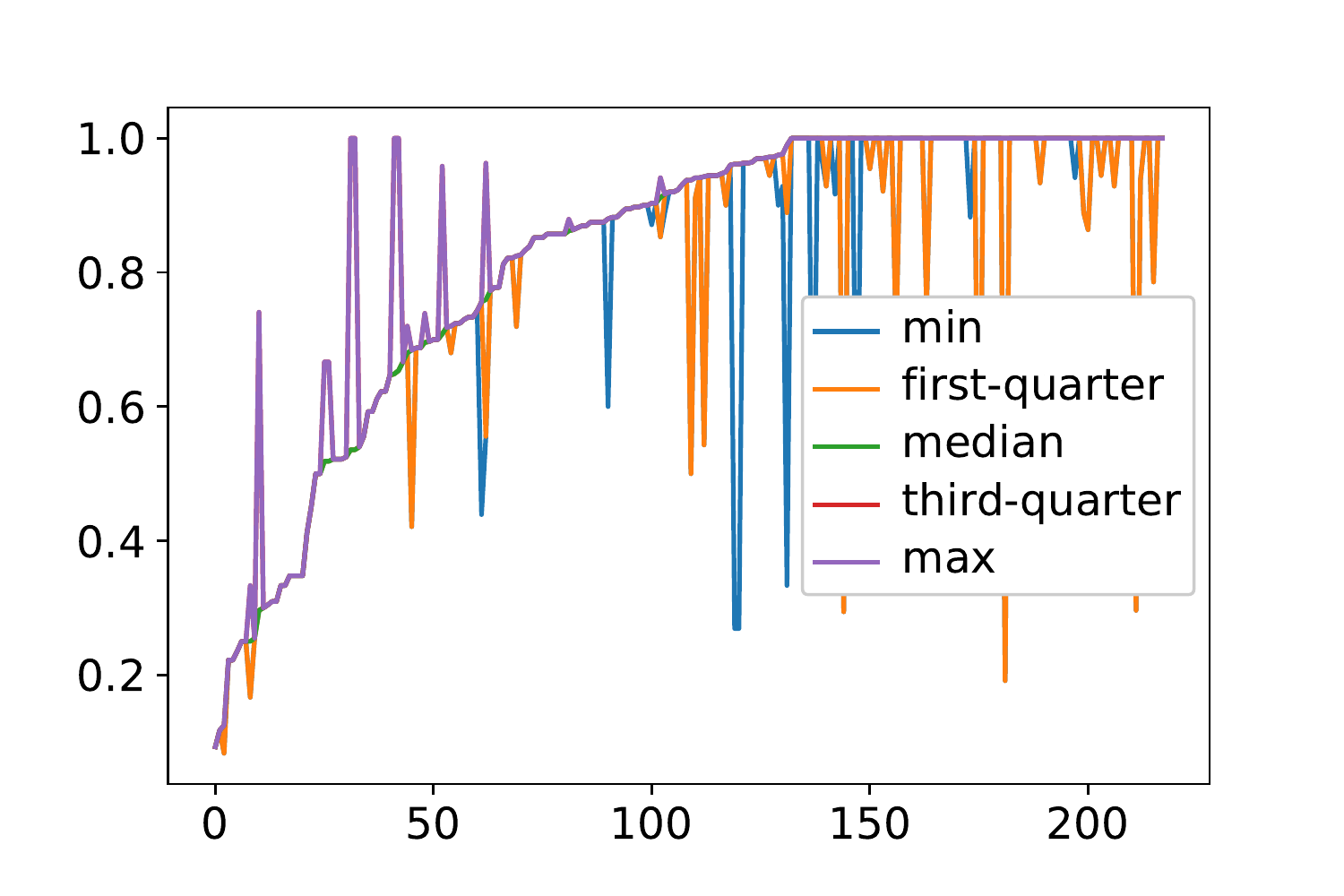}
	 \caption{Knee \%bug}
	 \label{fig:knee-b}
  \end{subfigure}
%  \hspace{-0.1in}
  \begin{subfigure}{0.24\textwidth}
    \includegraphics[width=4.8cm]{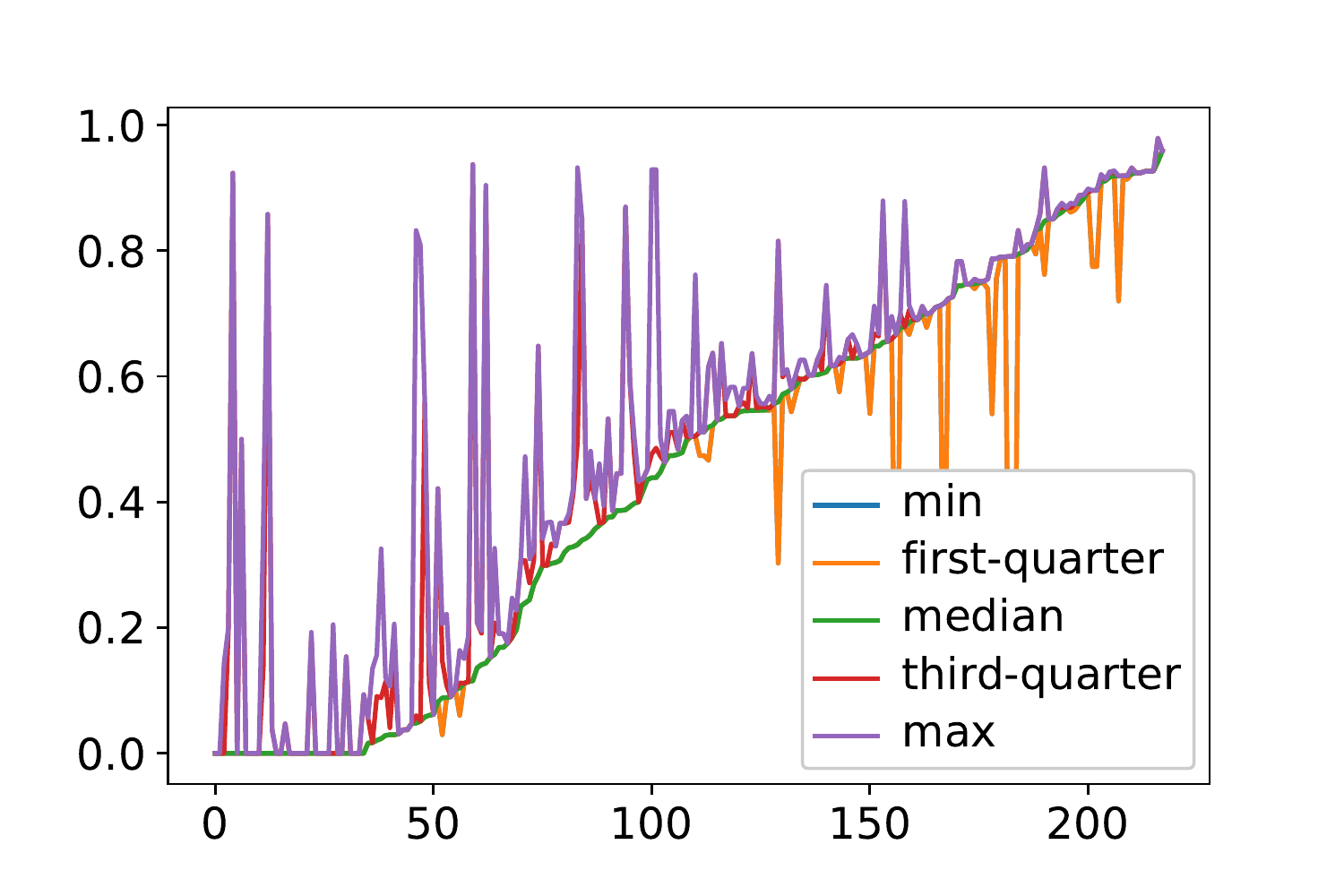}
	\caption{Knee \%reducedCost}
   \label{fig:knee-c}
  \end{subfigure}
  \begin{subfigure}{0.24\textwidth}
    \includegraphics[width=4.8cm]{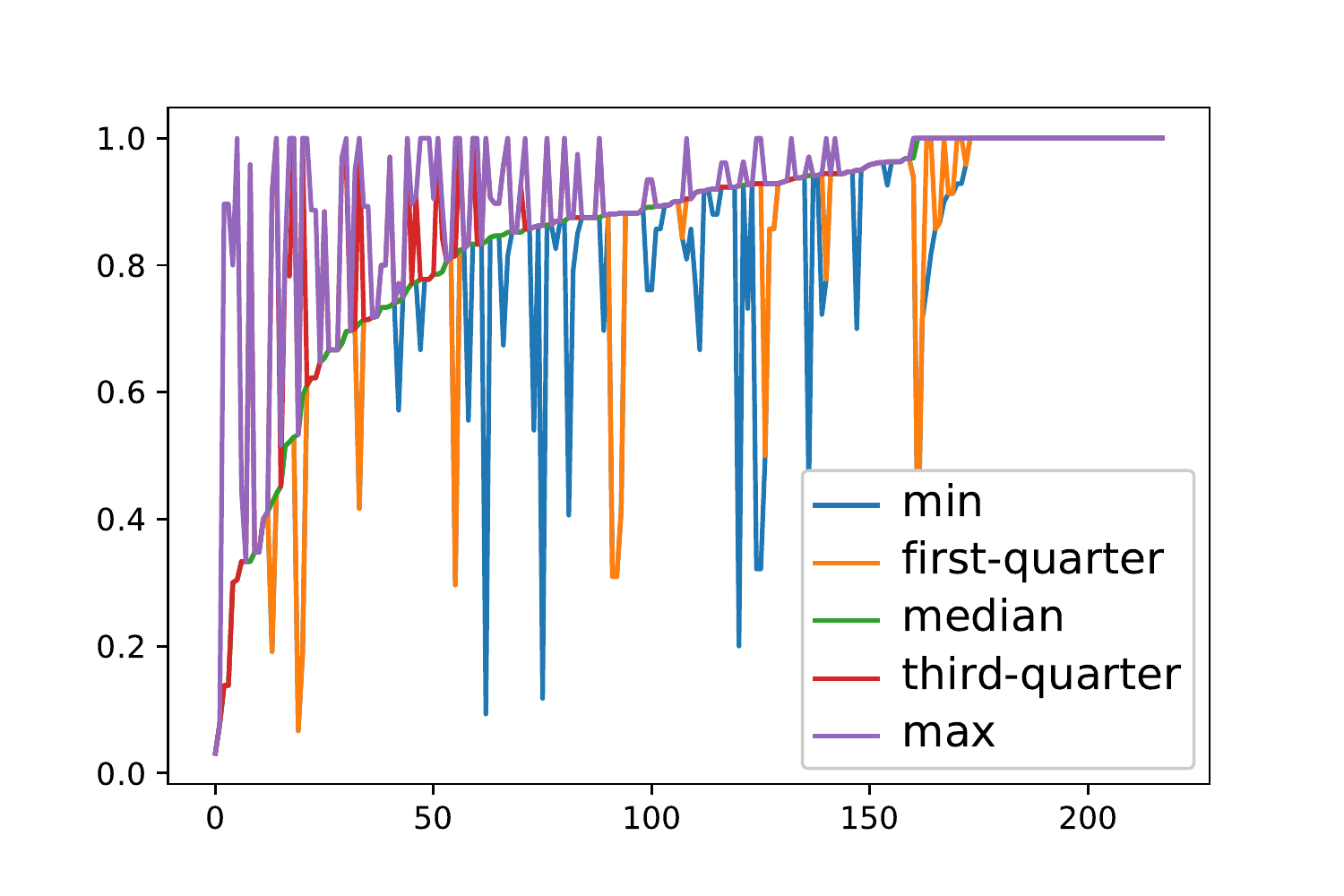}
	\caption{M0 \%bug}
    \label{fig:m0-b}
  \end{subfigure}
  \begin{subfigure}{0.24\textwidth}
    \includegraphics[width=4.8cm]{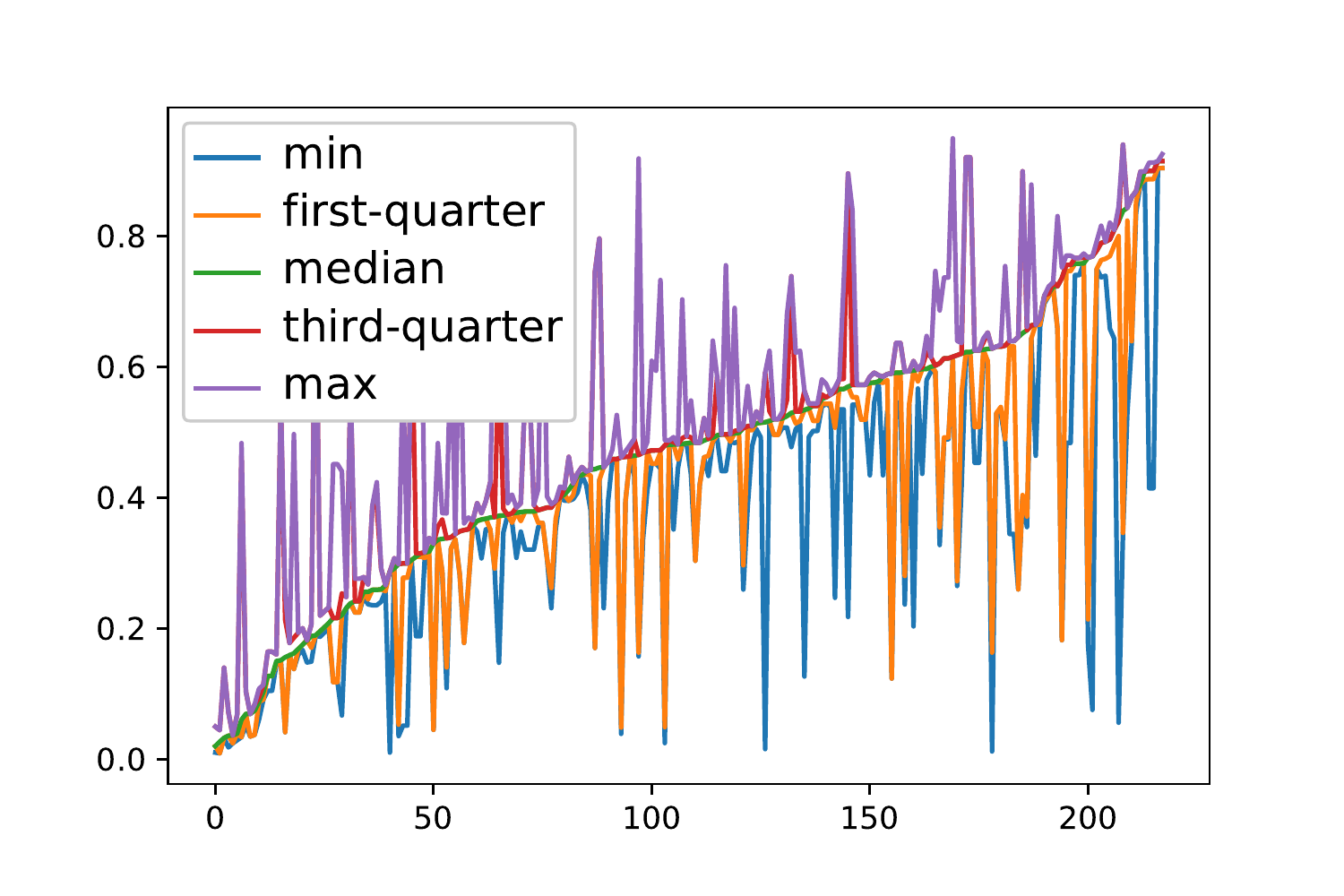}
	\caption{M0 \%reducedCost}
    \label{fig:m0-c}
  \end{subfigure}
  \caption{Stability of performance in terms of 1000 cross validations }
  \label{fig:random}
\vspace{-0.1in}
\end{figure*}
\subsection{Stability of Performance in Terms of 1000 Cross Validations}
\label{subsec:dis_stability}

We have conducted 1000 cross validations to tune the optimal parameter on the randomly-selected 2/3 training tasks and test the tuned parameter on the remaining tasks (see Section \ref{subsec:design_setup}).
One may want to know, for each experimental task, whether the performance is stable across all validations (i.e., under different tuned parameter values).
Figure \ref{fig:random} shows the min, first-quarter, median, third-quarter, and max value of \textit{\%bug} and \textit{\%reducedCost} for each experimental task (order by the median performance).
Due to space limit, we only present the results for the best four methods (i.e., \textit{Trend}, \textit{Peak}, \textit{Knee}, and \textit{M0}), and present other results on our website.

From Figure \ref{fig:random}, we can see that the performance of \textit{Trend} is more stable than other three methods in terms of 1000 cross validations. 
This again indicates the effectiveness of \textit{Trend}.

For \textit{Trend}, we can see that in 93\% (204/218) experimental tasks, the min and first-quarter of \textit{\%bug} obtained by \textit{Trend} is the same as the median \textit{\%bug}.
In 91\% (199/218) experimental tasks, the median \textit{\%bug} is the same as the max \textit{\%bug}.
For \textit{\%reducedCost}, in 87\% (191/218) experimental tasks, the min and first-quarter performance is the same as the median performance. 
% in 87\% (191/218) experimental tasks, the min and first-quarter performance is the same as the median performance. 
That is to say, for most experimental tasks, under different parameter values tuned by different training dataset, the performance remains almost unchanged. 
This implies the stability of \textit{Trend} in terms of the 1000 cross validations, and further proves its effectiveness.
We also noticed that for \textit{Trend} method, \textit{\%bug} is more stable than \textit{\%reduceCost}. 
This is what the project managers expect.
Because they mentioned the premises for making the crowdtesting more cost-effective is that a sufficient number of bugs should be detected.

We also examine the crowdtesting tasks whose min \textit{\%bug} is much smaller than its median \textit{\%bug} when using \textit{Trend}.
Results turn out that these projects have 18 successive reports during when the number of bugs remains unchanged and is far fewer than the total number of bugs (like the Rise-Stay-Rise pattern in Section \ref{subsec:background_observations}). 
When the tuned parameter value is 18 based on the training set, the predicted close point is right located in the end of the 18 successive reports, therefore the \textit{\%bug} is low. 
When the tuned parameter value is bigger than 18, this would not happen. 
This is why the method is not stable enough when applied on these several projects.

\begin{figure}[t!]
\centering
\includegraphics[width=8cm]{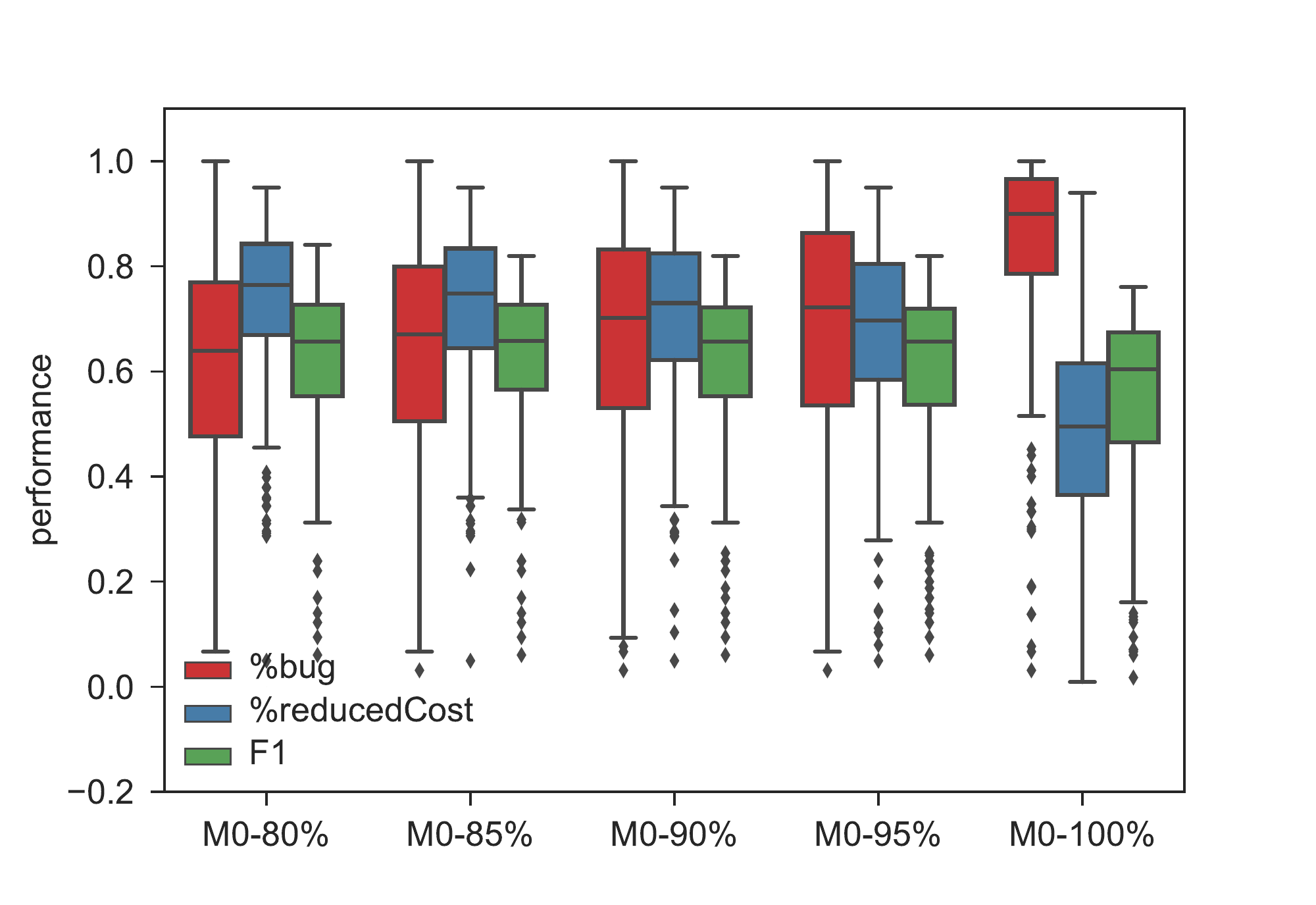}
\caption{Performance of M0 under customization}
\label{fig:advantageM0}
% \vspace{-0.1in}
\end{figure}

\subsection{Advantage of \textit{M0} Method}
\label{subsec:dis_advantage}

We have mentioned that \textit{Trend} and \textit{M0} are the best two methods, with \textit{M0} slightly inferior than \textit{Trend}.
% We have mentioned that \textit{M0}, which is based on capture-recapture model, is as effective as the best method \textit{Trend}, and only a little inferior in stability.
Nevertheless, method \textit{M0} provides additional flexibility over \textit{Trend}, i.e. customization of the close point according to user's preference.
In detail, the experimental results in Section \ref{sec:result} assumes all the bugs should be reported.
If the users hope to further save the cost, he can customize the method to close the task as long as \textit{K\%} (e.g., 80\%) of bugs are reported.

Figure \ref{fig:advantageM0} shows the performance of \textit{M0} when the customized {K\%} is \textit{80\%, 85\%, 90\%, 95\%, 100\%} (i.e., the default \textit{M0}).
We can easily see that a lower \textit{K\%} can save more cost with fewer detected bugs.
For example, default \textit{M0} can detect 91\% bugs with 49\% reduced cost.
When we customize \textit{K\%} as 80\%, a median of 65\% bugs is reported with 77\% reduced cost.

The reason why \textit{M0} can customize the close point is because it can obtain the estimated total number of bugs during the crowdtesting process.
With the estimated total number of bugs, and the number of detected bugs so far, we can know how much percentage of bugs have been reported.
However, we noticed that the total number of bugs are usually underestimated (i.e., customizing \textit{K\%} as 80\%, only a median of 65\% bugs found), we will explore this in future work. 
% Taken in this sense, \textit{Peak}, as well as other four CRC-based methods can also be customized. Due to space limit, we do not present the results.

\subsection{Threats to Validity}
\label{subsec:dis_threats}

The external threats concern the generality of this study.
Firstly, our experiment data consists of 218 crowdtesting tasks collected from one of the Chinese largest crowdsourced testing platforms.
We can not assume that the results of our study could generalize beyond this environment in which it was conducted.
However, the diversity of tasks and size of dataset relatively reduce this risk. 
% Even though the data might be different, the methods presented in this paper are largely applicable.
% and potential extension methods can tune parameter values easily on a new dataset.
Secondly, our designed methods are largely dependent on the report's attributes (i.e., whether it contains a bug; and whether it is the duplicates of previous ones) assigned by the task requesters.
This is addressed to some extent due to the fact that we collected the data after the crowdtesting tasks were closed, and they have no knowledge about this study to artificially modify their assignment.

Internal validity of this study mainly questions the representativeness of the 8 designed methods.
We have surveyed related work about quality insurance, software inspection and review, software reliability, and the designed methods involve the bug detection trend, the defect arrival model, the bug trend curve, the capture-recapture model, etc. Therefore we believe that this set coveres a large variety of of existing work, and captures the representative dynamic defect prediction models applicable in crowdtesting context. 

Construct validity of this study mainly concerns the experimental setup for investigating the effectiveness of each method. 
We use the most frequent tuned optimal parameter values to examine the effectiveness. 
We also present the stability of the method in terms of 1000 cross validations to further prove its effectiveness. 
% To address this, we employed cross validation 1000 times to ensure the results reflect the stability of each method. We choose the specific cross validation with the median performance which we suppose can reflect the effectiveness of each method.  

% We rely on the duplicate labels of crowdsourced reports stored in repository to construct the ground truth.
% However, this is addressed to some extent due to the fact that testers in the company have no knowledge that this study will be performed for them to artificially modify their labeling.
% Besides, we have verified its validity through random sampling and relabeling.

\section{Related Work}
\label{sec:related}

In this section, we discuss two areas related to our work,
i.e., crowdsouced testing, software testing and reliability.

\subsection{Crowdtesting}
\label{subsec:related_crowdtesting}

Crowdtesting has been applied to facilitate many testing tasks, e.g.,
test case generation~\cite{chen2012puzzle}, usability testing~\cite{gomide2014affective}, software performance analysis~\cite{musson2013leveraing}, software bug detection and reproduction~\cite{maria2016reproducing}.
These studies leverage crowdtesting to solve the problems in traditional testing activities,
some other approaches focus on solving the new encountered problems in crowdtesting.

Feng et al. \cite{feng2015test,feng2016multi} proposed approaches to prioritize test reports in crowdtesting.
They designed strategies to dynamically select the most risky and diversified test report for inspection in each iteration.
Jiang et al. \cite{jiang2018fuzzy} proposed the test report fuzzy clustering framework by aggregating redundant and multi-bug crowdtesting reports into clusters to reduce the number of inspected test reports.
Wang et al.~\cite{wang2016towards,wang2016local,wang2017domain} proposed approaches to automatically classify crowdtesting reports.
Their approaches can overcome the different data distribution among different software domains, and attain good classification results.
Cui et al. \cite{cui2017who,cui2017multi} and Xie et al.~\cite{xie2017cocoon} proposed crowdworker selection approaches to recommend appropriate crowdworkers for specific crowdtesting tasks.
These approaches considered the crowdworker's experience, relevance with the task, diversity of testing context, etc., and recommend a set of workers who can detect more bugs.

In this work, we focus on predicting when to close a crowdtesting task, which is valuable to improve the cost-effectiveness of crowdtesting and not explored in existing work.

\subsection{Software Testing and Reliability}
\label{subsec:related_inspection}

Many existing approaches proposed risk-driven or value-based analysis to prioritize or select test cases \cite{wang2017qtep,shi2015comparing,harman2015empirical,saha2015information,henard2016comparing,panichella2015improving}, so as to improve the cost-effectiveness of testing.
However, none of these is applicable to the emerging crowd testing paradigm where task requesters typically have no control over online crowdworkers's dynamic behavior and uncertain performance.
% crowdworkers are encouraged to come and perform the testing tasks at any time and could not be prioritized.

There are several researches focusing on studying the time-series models for measuring software reliability and predicting when to stop testing and release a software product \cite{garg2011stop,garg2013method,iqbal2013software}.
These researches have proposed different types of software reliability models to estimate the reliability of a software component for quality control purpose. 
Among them, we adopt two most promising models (i.e., \textit{Rayleigh's defect arriving model} and \textit{knee model}) for the close prediction of crowdtesting.

Another body of previous researches aimed at optimizing software inspection, which also concerned predicting the total and remaining number of bugs.
Eick et al. \cite{eick1992estimating} reported the first work on employing capture-recapture models in software inspections to estimate the number of faults remaining in requirements and design artifacts.
Following that, several researches focused on evaluating the influence of number of inspectors, the number of actual defects, the dependency within inspectors, the learning style of individual inspectors, on the capture-recapture estimators' accuracy \cite{briand2000comprehensive,walia2009evaluating,chun2006estimating,rong2017towards,mandala2012application,goswami2015using,vitharana2017defect}.
The aforementioned approaches are based on different types of capture-recapture models, and results turned out \textit{MhJK}, \textit{MhCH}, and \textit{MtCH} are the most effective estimators.
We have reused all these estimators and experimentally evaluated them.

% Briand et al. \cite{briand2000comprehensive} and Walia et al. \cite{walia2009evaluating} evaluated the influence of number of inspectors, the number of actual defects, the dependency within inspectors on the capture-recapture estimators' accuracy.
% Chun \cite{chun2006estimating} and Rong et al. \cite{rong2017towards} investigated the influence of dependency within inspectors on the estimation accuracy.
% Mandala et al. \cite{mandala2012application} investigated the cost and benefits of software inspections and predicting whether to schedule an inspection.
% % , and further analyzed the effect of team size on the estimation.
% Goswami et al. \cite{goswami2015using} investigated the influence of learning style of individual inspectors on the defect detection effectiveness.
% Vitharana et al. \cite{vitharana2017defect} investigated the extent of defect propagation at the project-level during early lifecycle phases, and its influence on software inspection efficiency.

\section{Conclusion}
\label{sec:conclusion}

It is valuable to automatically decide when to close a crowdtesting task so as to improve the cost-effectiveness of crowdtesting.
This paper first investigates the necessity and feasibility for close prediction of crowdtesting tasks.
Then it designs 8 methods to conduct the close prediction, respectively based on the bug trend, bug arrival model, and capture-recapture model.
Evaluations are based on 218 crowdtesting tasks from one of the largest crowdtesting platforms in China, and
results show that a median of 91\% bugs can be detected with 49\% reduced cost.

This paper also provides a cautionary tale that verbatim reuse of methods from other fields  may  not  produce  the  best  results  of  crowdtesting.
Specifically, we show the capture-recapture models from software inspections do not work well on crowdtesting data.
Furthermore, the most straightforward method (i.e., \textit{Trend}) can produce the most effective performance in close prediction of crowdtesting tasks.

It should be pointed out that the presented material is just the starting point of the work in progress. We are closely collaborating with {\company} crowdtesting platform and begin to deploy the \textit{Trend} and \textit{M0} method online. 
Returned results will further validate the effectiveness, as well as guide us in improving the methods.

%\section{Acknowledgments}
%This work is supported by the National Natural Science Foundation of China under grant No.61602450, No.6143200, and China Scholarship Council.
%We would like to thank the
%testers in Baidu for their great efforts in supporting this work.

%% Bibliography
%\bibliography{bibfile}

% \bibliographystyle{ACM-Reference-Format}
%\balance
% \bibliographystyle{ACM-Reference-Format}
% % \balance
% \bibliography{paper}

\bibliographystyle{plain}
% \bibliography{sigproc} % don’t import bib

               % import bbl

%% Appendix
%\appendix
%\section{Appendix}
%
%Text of appendix \ldots

\end{document}